\documentclass[final,3p,times,twocolumn]{elsarticle}

\usepackage{graphicx}
\usepackage{epsfig}
\usepackage{epstopdf}
\usepackage{url}
\usepackage{amssymb}
\usepackage{amsmath}
\usepackage{multirow}
\usepackage{verbatim}
\usepackage{subfigure}

\usepackage{lineno}

\journal{Nuclear Instruments and Methods A}

\begin{document}

\begin{frontmatter}



\title{Optical Transition Radiation Monitor for the T2K Experiment}

\author[york]{S. Bhadra} 
\author[uoft]{M. Cadabeschi}
\author[uoft]{P. de Perio}
\author[york]{V. Galymov}
\author[uoft,york]{M. Hartz\corref{cor1}}
\ead{mhartz@physics.utoronto.ca}
\cortext[cor1]{Corresponding author}
\author[york]{B. Kirby\fnref{ubc}}
\author[triumf]{A. Konaka}
\author[uoft]{A.D. Marino\fnref{cu}}
\author[uoft]{J.F. Martin\fnref{ipp}}
\author[triumf]{D. Morris}
\author[york]{L. Stawnyczy}

\address[uoft]{University of Toronto, Department of Physics, Toronto, Ontario, Canada}
\address[triumf]{TRIUMF, Vancouver, British Columbia, Canada}
\address[york]{York University, Department of Physics and Astronomy, Toronto, Ontario, Canada}
\fntext[ubc]{Now at University of British Columbia, Department of Physics and Astronomy, Vancouver, British Columbia, Canada}
\fntext[cu]{Now at University of Colorado at Boulder, Department of Physics, Boulder, Colorado, U.S.A.}
\fntext[ipp]{Also at Institute of Particle Physics, Canada}

\begin{abstract}
An Optical Transition Radiation monitor has been developed for the proton beam-line of the T2K long base-line neutrino oscillation experiment. The monitor operates in the highly radioactive environment in proximity to the T2K target. It uses optical transition radiation, the light emitted from a thin metallic foil when the charged beam passes through it, to form a 2D image of a 30 GeV proton beam. One of its key features is an optical system capable of transporting the light over a large distance out of the harsh environment near the target to a lower radiation area where it is possible to operate a camera to capture this light. The monitor measures the proton beam position and width with a precision of better than 500~$\mu$m, meeting the physics requirements of the T2K experiment.

\end{abstract}


\end{frontmatter}


\section{Introduction}
\label{sec:intro}
The properties of neutrinos continue to puzzle and challenge scientists in spite of tremendous progress on both the theoretical and experimental fronts.
T2K (Tokai-to-Kamioka) is a long-baseline neutrino experiment \cite{t2k_nim} searching for neutrino flavour changes in a neutrino beam 
that is generated at the Japan Proton Accelerator Research Complex (J-PARC).  The main goal is to measure the last unknown lepton sector mixing angle $\theta_{13}$ by observing $\nu_{e}$ appearance from a $\nu_{\mu}$ beam. In addition, from the disappearance measurement $\nu_{\mu} \rightarrow \nu_{\tau}$ oscillation parameters $\Delta m^2_{32}$ and $\sin^2{2\theta_{23}}$ will be measured with a precision of $\delta(\Delta m^2_{32})\mathord{\sim}10^{-4} \mbox{ eV}^2$ and $\delta(\sin^2{2\theta_{23}})\mathord{\sim} 0.01$. The neutrino beam at J-PARC is initiated by 30 GeV protons striking a graphite target resulting in many secondary particles, especially pions and kaons.  After being focussed by magnetic horns, the short-lived pions and kaons are allowed to decay freely inside a 100~m long helium filled region.  The resulting neutrino beam consists predominantly of muon neutrinos, with a very small component of electron neutrinos. 

Near detectors ND280 and INGRID are located on the J-PARC site 280 m from the target. The INGRID detector is placed on the axis defined by the proton beam direction and monitors the neutrino beam direction. ND280 is located $2.5^{\circ}$ off-axis to accept a narrow-band neutrino beam with peak energy around 600 MeV and measures the neutrino energy spectrum and interaction rates in the unoscillated state. The far detector, Super-Kamiokande (SK), 295 km away from J-PARC and also at $2.5^{\circ}$ off-axis,  studies changes in the beam after travel. 

\subsection{Motivation for the Optical Transition Radiation Monitor} 

\begin{figure}
\centering
\includegraphics[width=0.47\textwidth]{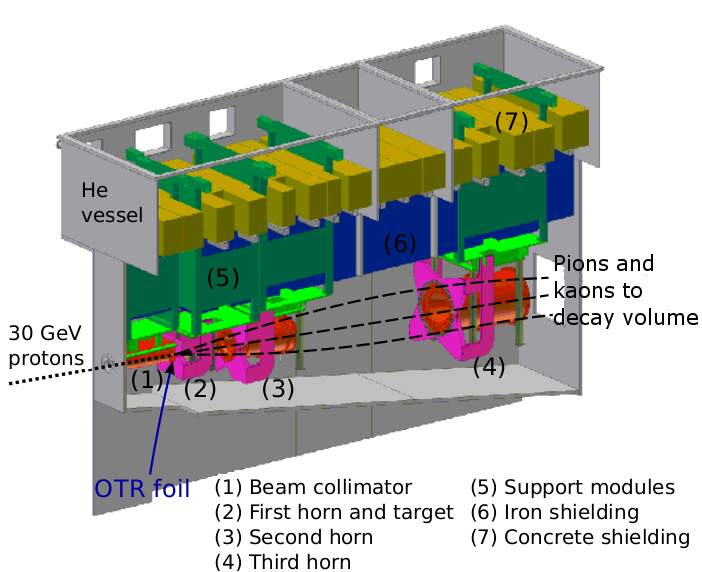}
\caption{This drawing shows the T2K target area.  The OTR foil is placed between the beam collimator and the first horn.}
\label{fig:targetarea}
\end{figure}

A precision measurement of the oscillation parameters depends on an accurate determination of the position, profile and angle of the proton beam which produces neutrinos. Due to the point-to-parallel nature of the horn focussing of the produced mesons, the off-axis angle and therefore the neutrino spectrum in the SK detector is affected by the primary proton beam position and angle. The position and direction of the proton beam needs to be measured at the target with a precision of 1 mm and 0.5 mrad, respectively, in order to ensure that the contribution from this effect to the systematic errors of the T2K physics measurements is small. The beam size must be measured with a precision of about 10\% to keep within the limits required for radiation losses and target protection. An Optical Transition Radiation (OTR) monitor measures the beam profile and position just upstream of the target to the required precision and is the subject of this paper. 

Transition radiation is produced when charged particles traverse a boundary between materials with different dielectric constants. This was first theoretically predicted by Ginsburg and Franck \cite{ginsburg} and experimentally verified for vacuum-metal boundaries by Goldsmith and Jelley \cite{goldsmith}. Since then this technique has been used at a number of accelerators for measurements of beam characteristics \cite{fnal_otr}, \cite{cern_otr}, \cite{jparc_otr} where in general a thin foil is introduced in the path of the beam and the resulting OTR light is recorded with a camera.

For the T2K experiment, the expected dose near the target for 750kW operation is $5.4\times10^8$ Sv/hr, while the dose even at a distance 1 m perpendicular to the beam direction is $0.8\times10^4$ Sv/hr. This precludes the placement of electronics nearby. A long optical path is needed to transport OTR light out of the harsh radiation environment near the target to one more suitable for camera placement. 

Fig.~\ref{fig:targetarea} shows a side view of the T2K target area. A beam collimator, three focusing horns, the target (inside the first horn), concrete and iron shielding and the decay volume are all enclosed in a large vessel filled with helium. The collimator, target and horns are mounted under large, steel-framed support modules. The support modules and the shielding sit on mounts attached to the sides of the helium vessel. A series of mirrors focuses and transports the OTR light through channels in the iron and concrete  shielding, through a fused silica window in the helium vessel lid,  to a camera.  The camera and readout electronics sit outside of the target vessel in an environment that has a radiation dose 5 orders of magnitude smaller than the area near the first mirror. 

\section{Transition Radiation}

When a charged particle travels between two different media, the fields of the charged particle induce a polarization on the surface of the new medium, and they combine coherently to form transition radiation.  As discussed in \cite{jackson}, the formation depth $D$ for transition radiation is of the order of
\begin{linenomath*}
\begin{equation}
D=\frac{\gamma c}{\omega_{p}},
\end{equation}
\end{linenomath*}
where $\omega _{p}$, the plasma frequency of the medium (with an electron number density $n_{e}$), is defined as
\begin{linenomath*}
\begin{equation}
\omega_{p}^{2} =\frac{4\pi n_{e}e^{2}}{m_{e}}.
\end{equation}
\end{linenomath*}
For example, in solid titanium ($\omega_{p} \approx 1.34\times10^{16} \mbox{ s}^{-1}$), with a 30~GeV proton ($\gamma \approx $ 32), $D$ is of the order of 1~$\mu$m. Therefore only a thin layer of material is needed to produce transition radiation. 

The general expression for the number of photons, $N$, that are emitted in a frequency range  $d \omega$, into  a solid angle $d \Omega$,  when a charged particle passes at normal incidence from material 1 (with dielectric constant $\epsilon_{1} $) to material 2 (with dielectric constant $\epsilon_{2} $) is~\cite{ter-mikaelian} 
\begin{linenomath*}
\begin{multline}
\frac{d^{2}N}{d\omega d\Omega}= \frac{2e^{2}\beta^{2}\sqrt{\epsilon_{2}}sin^{2}\theta cos^{2}\theta}{\pi h c \omega}  \times    \left| \frac{1}{(1-\beta^{2}\epsilon_{2}cos^{2}\theta)} \right|^{2} \times  \\    
  \left| \frac{(\epsilon_{1}-\epsilon_{2})(1-\beta^{2}\epsilon_{2}-\beta\sqrt{\epsilon_{1}-\epsilon_{2}sin^{2}\theta}) }{(1-\beta\sqrt{\epsilon_{1}-\epsilon_{2}sin^{2}\theta})\left (\epsilon_{1}cos\theta+\sqrt{\epsilon_{1}\epsilon_{2}-\epsilon_{2}^{2}sin^{2}\theta}\right)} \right| ^{2},
\label{eqn:genotr}
\end{multline}
\end{linenomath*}
where $\theta $ is the angle between the particle's velocity and the photon vector.  For the case of a relativistic charged particle ($\beta \sim 1$) moving from a material with $|\epsilon_{1}| > 1$,  into vacuum ($\epsilon_{2}=1$), this reduces to~\cite{otr_gitter}
\begin{linenomath*}
\begin{equation}
\frac{d^{2}N}{d\omega d\Omega}=\frac{2e^{2}}{\pi hc \omega}\times \frac{\theta^{2}}{(\theta^{2}+\gamma^{-2})^{2}}.
\label{eqn:otrforward}
\end{equation}
\end{linenomath*}
Note that the light is emitted in a narrow forward cone, with the maximum of the angular distribution at $\theta \sim 1/\gamma$.  For the case of a particle travelling through a thin foil, OTR is also produced in the backward direction when the particle enters the foil from vacuum. In this case, $\epsilon_{1}=1$ and $\epsilon_{2}=\epsilon$.  Equation~\ref{eqn:genotr}, with $\beta \sim 1$, now reduces to
\begin{linenomath*}
\begin{equation}
\frac{d^{2}N}{d\omega d\Omega}=\frac{2e^{2}}{\pi hc \omega}\times \left| \frac{\sqrt{\epsilon}-1}{\sqrt{\epsilon}+1} \right|^{2} \times\frac{\theta^{2}}{(\theta^{2}+\gamma^{-2})^{2}}.
\label{eqn:otrbackward}
\end{equation}
\end{linenomath*}

\begin{figure}[t]
\centering
\includegraphics[width=0.47\textwidth]{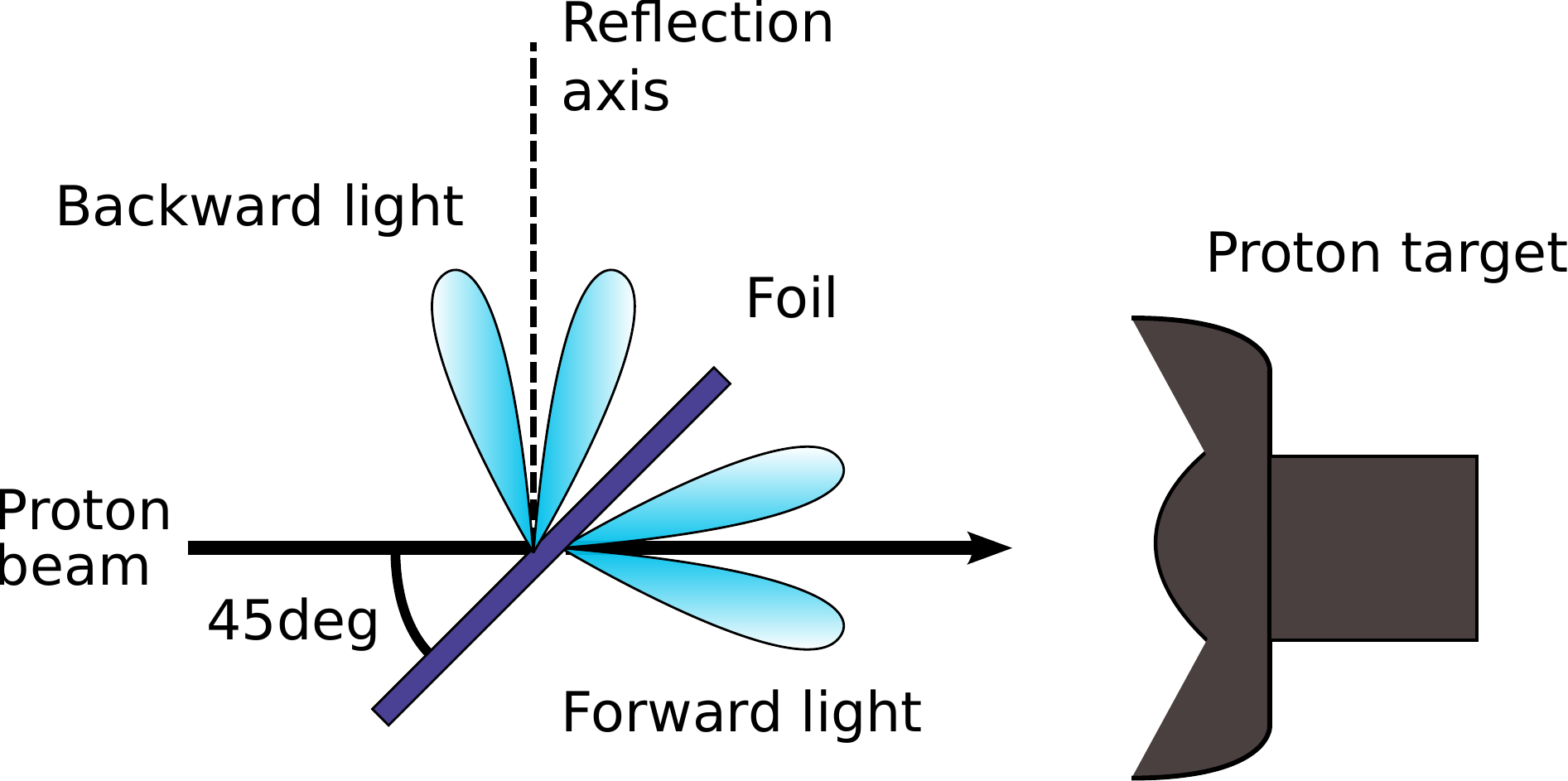}
\caption{This figures illustrates the direction of the forward and backward OTR light from a foil oriented at 45 degrees with respect to the beam.}
\label{fig:otrlobe}
\end{figure}

 The geometry changes slightly for the case where the particle is not at normal incidence to the surface.  The forward lobe, produced by the charged particle exiting the foil, is still oriented around the line of motion of the charged particle.  However, the backward, or reflected lobe, now surrounds the axis of reflection from the foil surface.  For a thin foil oriented at 45 degrees with respect to a beam of charged particles,  the backward lobe will be reflected at 90 degrees from the original beam direction, as shown in Fig.~\ref{fig:otrlobe}.  It is this backward lobe that we transport and detect in the T2K OTR monitor.  
 
\section{Optical System}
\label{sec:optics}

\subsection{Overview}

\begin{figure}[h]
\begin{center}
\includegraphics[width = 0.47\textwidth]{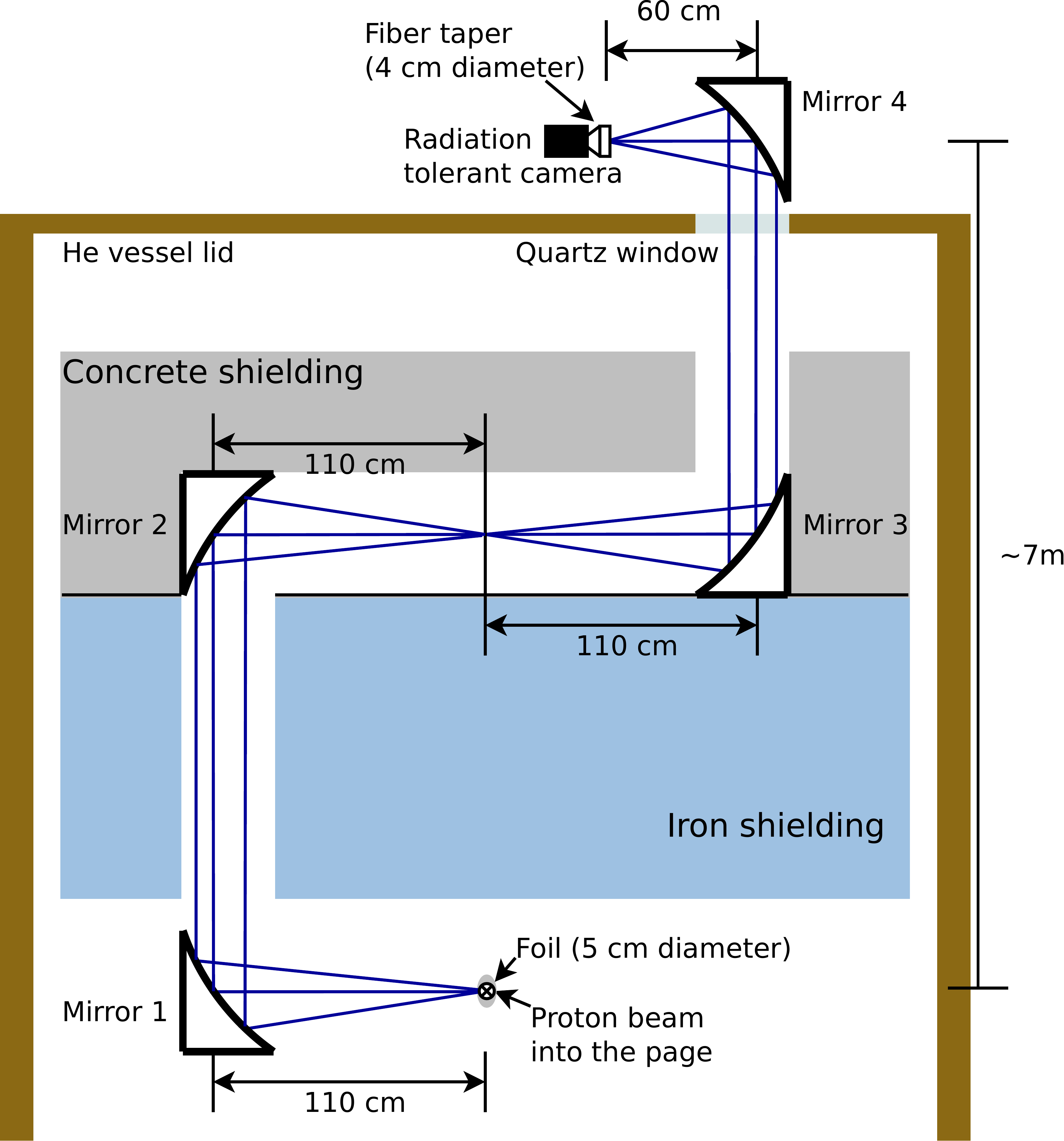} 
\caption{This figure shows a slice through the optical path of the OTR system where the proton beam is going into the page and striking the foil. Three light rays illustrate the focussing properties of the optics.}
\label{fig:otrpath}
\end{center}
\end{figure}

Fig.~\ref{fig:otrpath} shows the optical layout of the OTR system.  The beam strikes a foil oriented at 45 degrees with respect to the beam line.  As described in the previous section, the backward lobe of the OTR is emitted around the reflection axis, in this case 90 degrees relative to the incident proton beam direction.   The foil sits immediately upstream of the target, and is downstream of the beam collimator.  


The OTR light must travel through several bends in the shielding to avoid a direct path for the radiation from the target region.   A series of 4 parabolic mirrors transport the light through this path. These parabolic mirrors are 90 degrees off-axis, giving an effective focal length (the distance from the centre of the mirror to the focal point) twice the focal distance of the parent parabolic surface.  The light diverges from the foil to mirror 1, travels as a parallel beam to mirror 2 and comes to an intermediate focus halfway between mirrors 2 and 3 (see Fig.~\ref{fig:otrpath}). This pattern repeats using mirrors 3 and 4 before the final focus at the camera position.  A 25~cm diameter, fused silica window in the aluminum lid of the helium vessel allows the OTR light to emerge for capture at the camera situated on top of the lid. 

It is desirable to have the first mirror as far away as possible from the foil to reduce radiation exposure, and also the aperture size must be large enough to collect a large fraction of the light. The mirror diameter is limited by the maximum allowable size for the channels in the shielding. Given these considerations, mirror 1  is placed 110 cm from the foil (requiring an effective focal length of 110~cm) and has a diameter of 12 ~cm.  Mirrors 2 and 3 have the same focal length and size as mirror 1.  However mirror 4 has a shorter focal length of 30~cm (effective focal length of 60~cm).  This reduces the size of the foil image at the camera by 45\% to allow the image of the 5~cm diameter foil to fit within a 4~cm diameter fiber taper connected to the face of the camera.  The fiber taper reduces the image diameter from 4~cm to 1.1~cm to fit onto the size of the camera sensor. 

\subsection{Mirrors}

 Since there are four mirrors in the system, a mirror surface with high reflectivity is required to reduce light loss.  The parabolic mirrors used in this system were fabricated by B-Con Engineering Inc. out of solid aluminum  and coated with a uniform 400~nm thick layer of Al$_{2}$O$_{3}$, which has a reflectivity close to 100\%.  A small test mirror with this coating was irradiated with a proton beam at TRIUMF for the equivalent radiation dose that is expected for 130 years of operation at the location of mirror 1.  No significant change in the reflective properties of the test mirror was observed.  This provided confidence in the long term reflective properties using this coating.

\subsection{Camera}
\label{subsec:camera}
 The radiation dose level at the camera is estimated to be $\sim$1~kGy/year
 at 750~kW beam power.  A charge injection device (CID) camera 
 (Thermo Fisher Scientific 8710D1M) is used as it is radiation tolerant up to
 10~kGy. 
 The sensor pixel matrix consists of $755\times484$ sites with pixel dimensions of 12.0~$\mu$m $\times$ 13.7~$\mu$m. An analog signal is sent from the camera over 50 m of shielded cable to the data acquisition system at the ground level of the
 target building.

\begin{figure}
\begin{center}
\includegraphics[width = 0.45\textwidth]{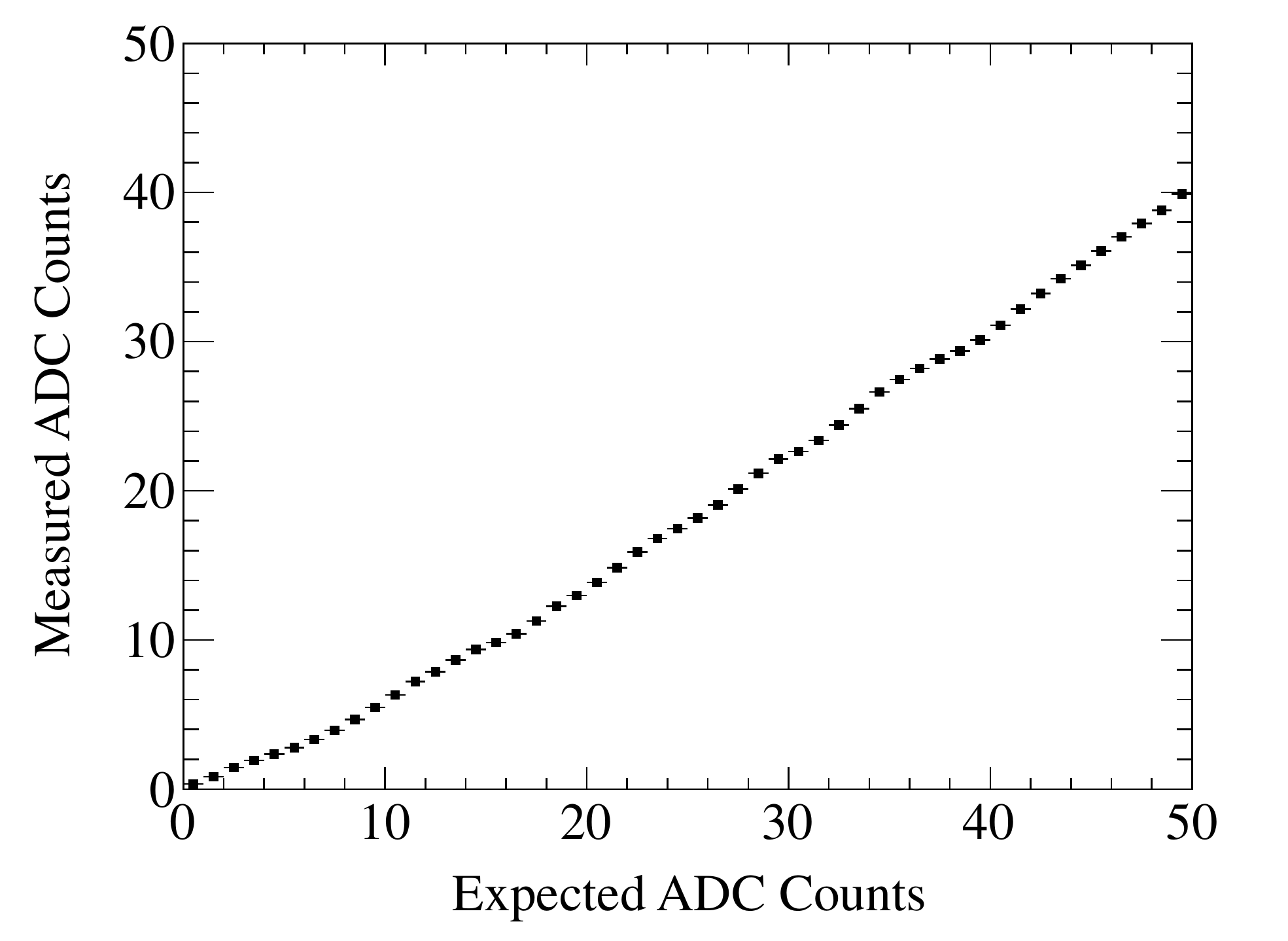}
\caption{The measured camera response in ADC counts is shown as a function of the expected 
ADC counts assuming linear operation. The response is shown here before the introduction of the ambient light and is averaged over the pixels.}
\label{fig:cameralinearity}
\end{center}
\end{figure}

During the initial operation of the neutrino beam-line, the low intensity of the delivered proton beam led to small OTR light yields. At these light levels the camera has a non-linear response as shown in Fig.~\ref{fig:cameralinearity}. This was understood to be caused by impurities in the silicon that trapped the collected charge. By adding uniform ambient light these traps could be populated and the response of the camera could be moved into the linear region. 

The camera outputs an interlaced video signal where even lines are read out $\sim17$ msec before odd lines.  The 
camera pixels collect charge at the arrival of the OTR light, immediately before the readout of the first even line. 
During the time between the charge collection and its readout, a pixel may lose charge through leakage currents. 
To study this effect, the ratio of neighboring even and odd pixels is taken for images of OTR light.  A constant
ratio of $0.81$ is observed, indicating an exponential decay of the charge with time constant $\tau=77$ msec.  
The measured ADC count for each pixel is corrected for this charge decay based on the readout time of the pixel.

\subsection{Predicted Photon Yield} 
\label{sec:photonyield}

The number of generated OTR photons for a material with dielectric constant $\epsilon$ can be estimated from Eq.~\ref{eqn:otrbackward}. Based on ray tracing simulations, the light collection efficiency for OTR light emitted at the centre of the foil with perfectly reflective mirrors is approximately 15\%.  Since light that reaches the camera reflects off four aluminum mirrors,  the generated OTR photon spectrum must also be multiplied by the wavelength dependent reflection coefficient for aluminum raised to the 4th power.  After mirror 4, the photons pass through the fiber taper, which has a transmittance of roughly 45\%, to strike the photocathode of the camera. 

A combining of all the acceptances and efficiencies (including the quantum efficiency of the camera sensor, which peaks at $~30$\% at 570~nm) yields the number of electrons recorded by the CID camera as a function of the wavelength of the initial OTR photons.  For a titanium alloy foil, integrating over the visible range results in $\sim 2.1\times10^{-5}$ electrons in the camera ($\sim 2.1\times10^{-8}$ ADC counts) per 30 GeV proton.     

During beam commissioning and early data taking periods, the number of protons per pulse varied from $1\times10^{11}$ to $1\times10^{13}$, equivalent to $0.13$ to $13$ ADC counts at the pixel sampling the peak of the light distribution.  At the lower end of this range, the OTR light yield from titanium is too low to image.  An aluminum foil gives $2-4$ times more light, depending on the surface roughness, due to higher reflectivity. The light yield from aluminum is, however, still insufficient for very low beam intensities, so a ceramic Al$_{2}$O$_{3}$:Cr$^{3+}$ $100$ $\mu$m thick wafer producing fluorescent light is used for very low intensity beam running.  This is a Demarquest AF995R wafer machined by the Niki Glass Company. The fluorescence is proportional to the energy deposited by the proton beam (apart from nonlinear effects discussed in Section~\ref{subsec:waferperform}). Accounting for optical and quantum efficiencies, the camera collects $6.0 \times10^{9}$ electrons per 30 GeV proton, sufficient to image the beam at very low intensities. A remotely controlled disk of neutral density filters in front of the camera can be rotated to optimize the light intensity for any given beam intensity, and this feature will be particularly useful as we move to high beam intensities where saturation at the camera can occur. 

\subsection{Prototype System}
\label{sec:prototype}

A prototype system was assembled to demonstrate the detector's ability to observe
transition radiation and measure the position and width of a particle
beam.
The prototype consisted of four parabolic mirrors with focal lengths and
relative distances scaled to 13.8\% of the full system size, test foils
held by a fixed support and a charge-coupled device (CCD) type photosensor
read out directly to a computer.
The system was tested in a NRC (Ottawa) electron linear accelerator capable of
producing a continuous electron beam with similar Lorentz factor $\gamma$
to the J-PARC proton beam, resulting in similar transition radiation
angular and spectral distributions.
Electron beam induced transition radiation was observed for test foils
composed of titanium-alloy, aluminum and graphite, with calibration holes
placed in the titanium-alloy foil also visible.
In order to identify potential background light sources the J-PARC target
station environment was simulated by enclosing the prototype in an
sealed bag filled with helium, at which point no additional
beam-induced light was observed.

The prototype system beam width and
 position measurement resolutions were estimated  to be 15\% and 0.2~mm, respectively, by
comparing the CCD camera results to the NRC wire detector measurements of the beam.  This provided confidence to
proceed to the design and construction of the full-scale system.

\section{Mechanical Design and Set-up}
\label{sec:mechanical}

\begin{figure*}
\begin{center}
\subfigure[The OTR components near the beam.]{
\includegraphics[width = 0.45\textwidth]{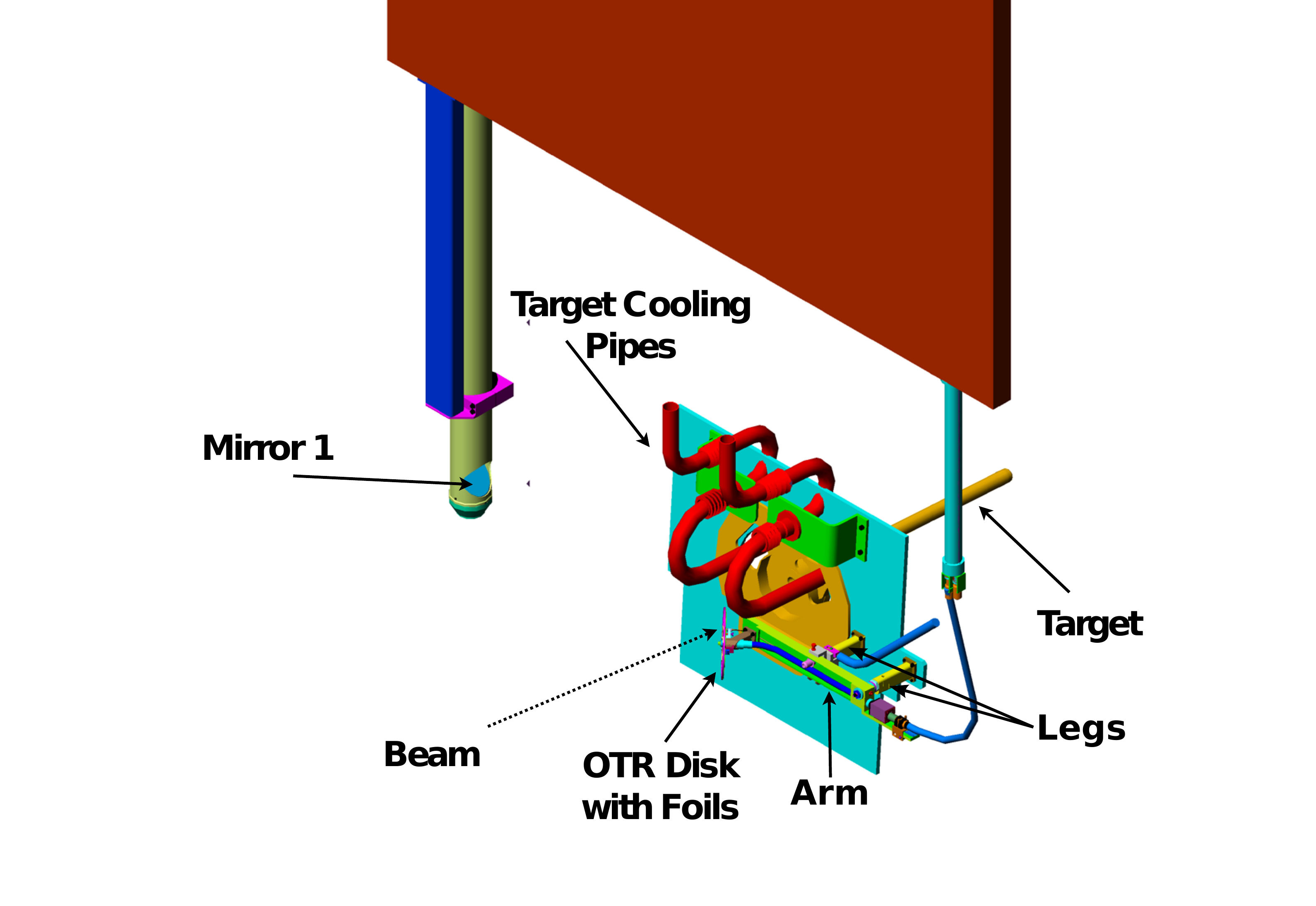}
\label{fig:OTR_upstream_schematic}}
\subfigure[A view of the OTR system from the rear, showing components mounted on the front plate of the horn support module.]{
\includegraphics[width = 0.45\textwidth]{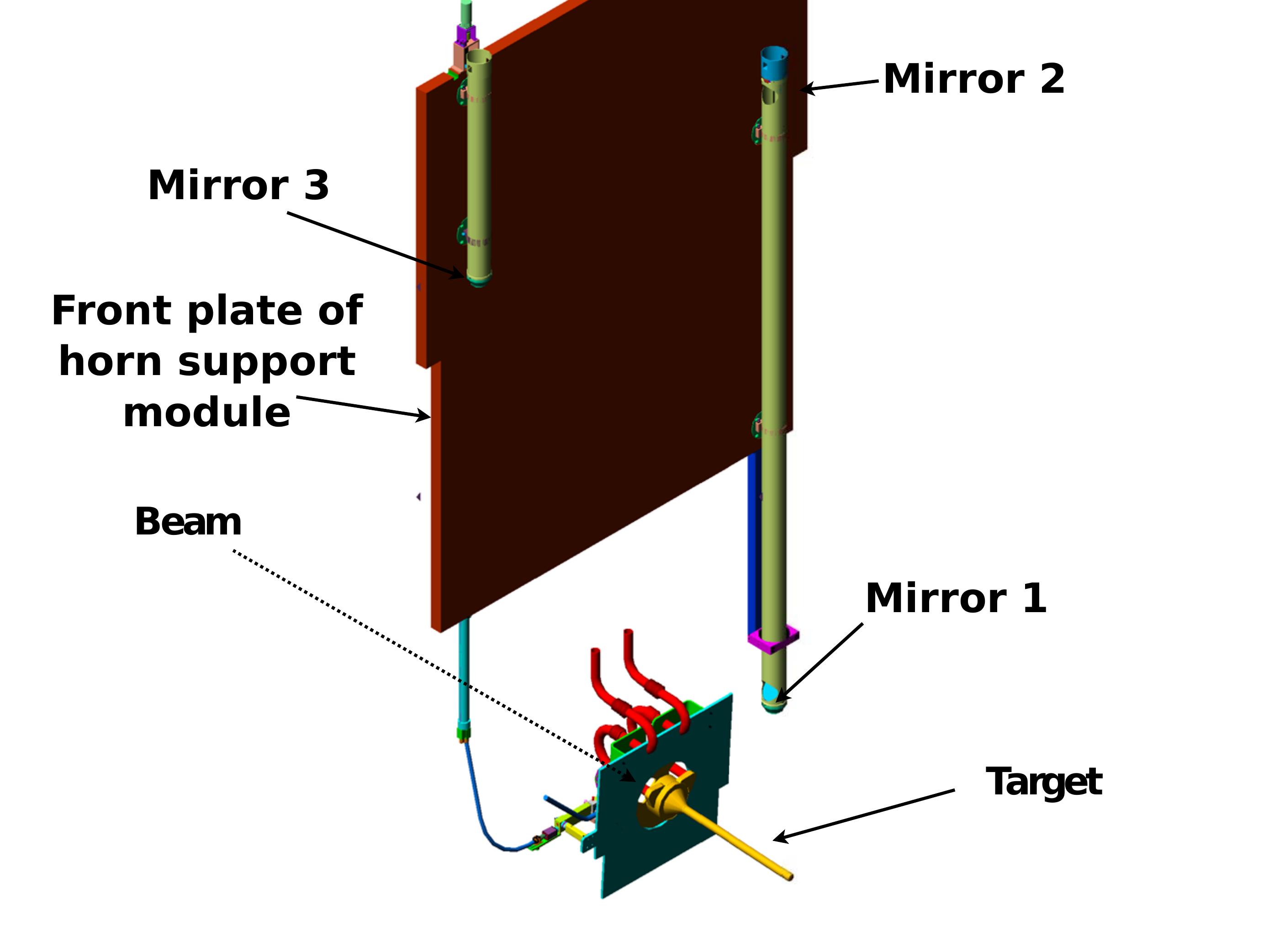}
\label{fig:OTR_downstream_schematic}}
\end{center}
\caption{The OTR system components}
\label{fig:otr_syst_model}
\end{figure*}

The OTR mechanical systems are shown schematically in Fig.~\ref{fig:OTR_upstream_schematic}~and~~\ref{fig:OTR_downstream_schematic}. They were designed to satisfy several requirements arising from the inaccessible high-radiation environment near the beam: 
\begin{list}{\labelitemi}{\itemsep=0pt\parsep=0pt}
\renewcommand{\labelitemi}{{\tiny$\bullet$}}
\item the ability to continually calibrate the optics
\item stability with temperature
\item long-term robustness 
\item ease of remote maintenance for part replacement.
\end{list} 

\subsection{The Foil Disk System}
\label{sec:foils}
   
\begin{figure}
\begin{center}
\includegraphics[width=0.38\textwidth]{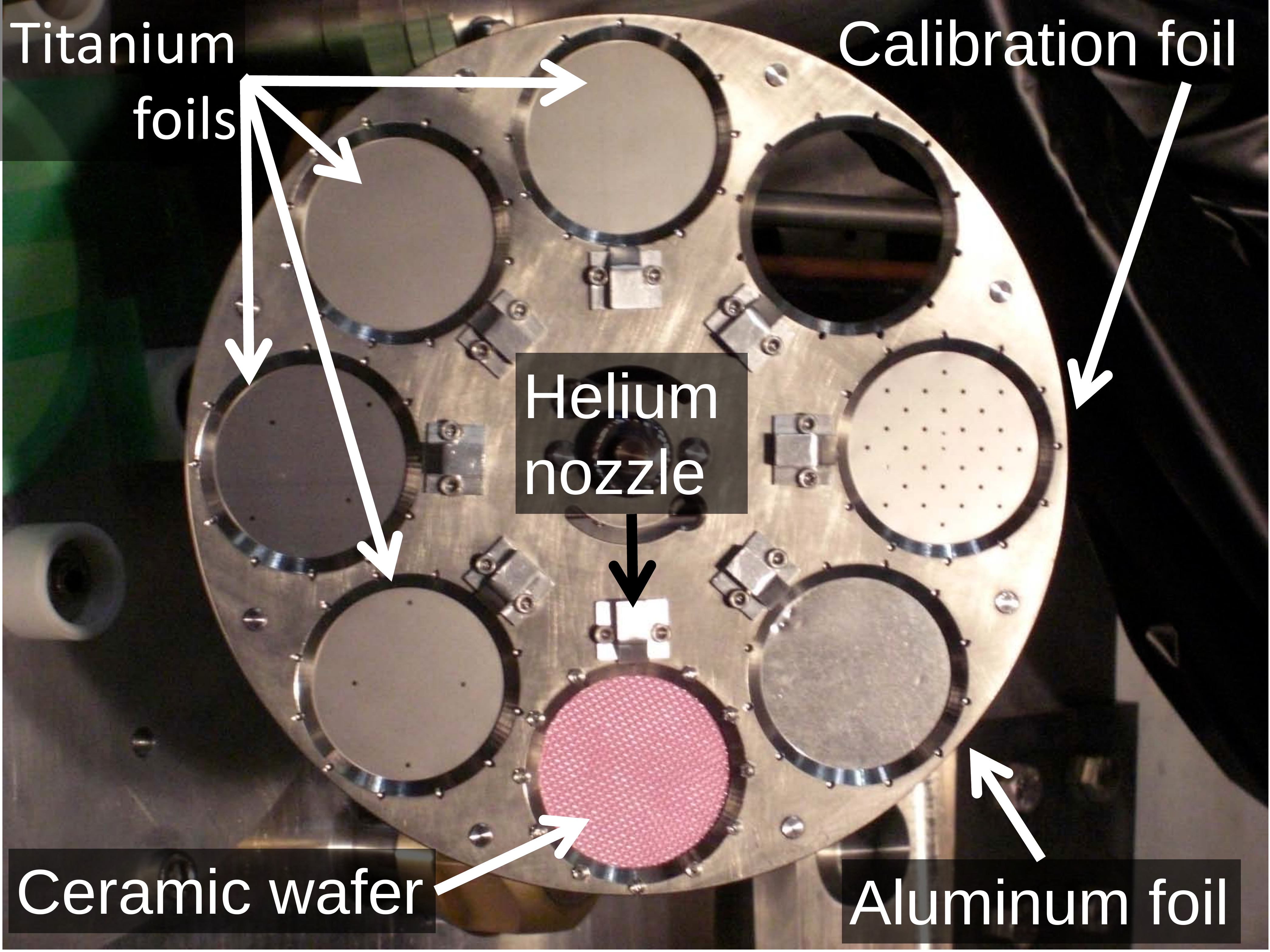}
\end{center}
\caption{The foil disk (upstream side).}
\label{fig:foil_disk}
\end{figure}

The OTR foils are mounted on a disk carousel which has eight hole positions of diameter 50~mm, as shown in Fig.~\ref{fig:foil_disk}. With the disk mounted at 45 degrees to the beam axis, the foils cover the full beam acceptance defined by the 30~mm diameter hole in the upstream collimator. Seven hole positions are occupied as described in Table~\ref{table:list_of_foils} and the eighth position is empty. Two positions are filled with a ceramic wafer and an aluminum foil, as discussed in Section~\ref{sec:photonyield}.
Having four titanium alloy foils allows for replacement in case of foil damage at high intensity.
A fifth titanium alloy foil has a pattern of precisely laser-machined holes and is used only for calibration when there is no beam.
Titanium was chosen after various foil materials were studied using the programs MARS \cite{MARS_1,MARS_2,MARS_3,MARS_4} and FemLab, in addition to approximate theoretical calculations. 
 Although temperature rise is not a problem for several materials, only titanium alloys have sufficient yield strength to survive the stresses of the full-intensity T2K beam, with a safety factor of about four.  Based on our studies, the titanium alloy 15V-3Cr-3Sn-3Al (the numbers refer to the percentages of the elements in the alloy composition) was chosen for the OTR foils.

\begin{table*}[ht]
\begin{center}
\caption{Foils used in the OTR system}
\label{table:list_of_foils}
\vspace{0.1in}
\begin{tabular}{l|l|l}
\hline
Material (number of foils) & Thickness ($\mu$m) & Operation \\
\hline
\hline
AF995R (1) & 100 &  $<1$ kW beam power \\ \cline{1-2} \cline{3-3}
Al 1100 (1) &   & $1-40$ kW beam power \\ \cline{1-1} \cline{3-3}
 & &        \\ 
Ti 15-3-3-3 (4) & 50 & $>8$  kW beam power\\ 
 & &   \\ 
 & &   \\ \cline{1-1} \cline{3-3}
Ti 15-3-3-3 (1) & & calibration with no beam\\ \hline
\end{tabular}
\end{center}
\end{table*}

\begin{figure}
\begin{center}
\includegraphics[width=0.4\textwidth]{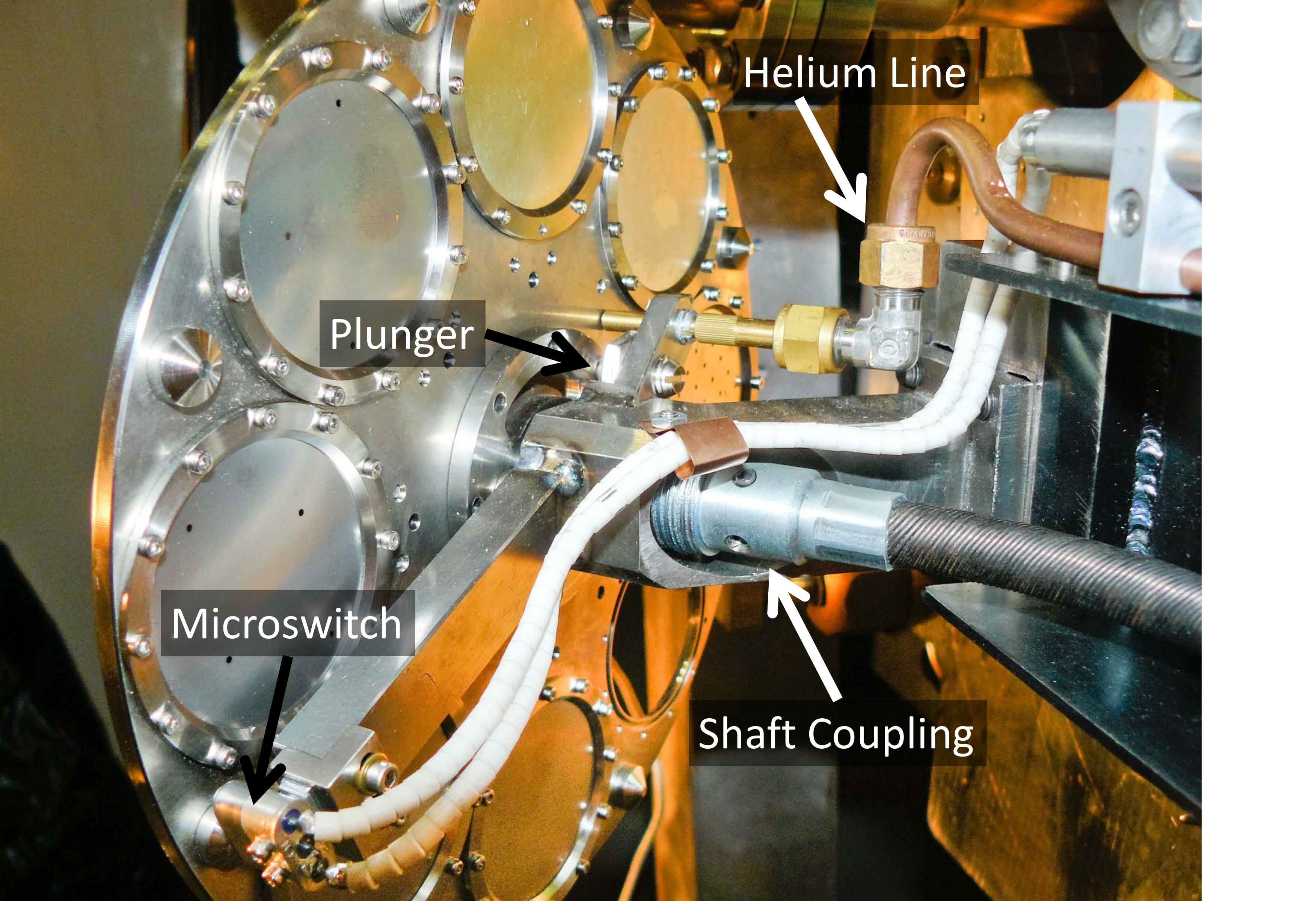}
\end{center}
\caption{The foil disk (downstream side).}
\label{fig:foil_disk_rear}
\end{figure}

Each metal foil is stretched by a clamping ring (see Fig.~\ref{fig:foil_disk_rear}) with a machined ridge which forces the foil edge into a corresponding circular groove on the disk. The size and shape of the ridge and groove were determined by iterative design and testing  in order to provide the required tension. Finite element analysis indicated that a stress of somewhat less than 100~MPa will result from the expected maximum intensity beam of $3.3\times 10^{14}$ protons per pulse. In order to keep the foils flat, a tension stress comfortably greater than this is required. The final clamping mechanism produces a measured stress of 190~MPa.

The disk is mounted on an ``arm'' held by two ``legs'' (see Fig.~\ref{fig:otr_syst_model}) that are attached precisely (dowel-pinned) to a large aluminum plate which is part of the target/horn assembly. 
The disk, arm and legs are made of titanium, chosen for its low coefficient of thermal expansion. Since the aluminum plate is actively cooled with helium but the arm is not, relative movement is thus minimized. 
The arm temperature is measured by a thermocouple. At the highest beam power reached so far, 145 kW, the temperature rise is $8^{\circ}$C.
 
Any of the foils can be positioned in the beam by rotating the disk.  It is important that the foil is centered on the horn axis so that the beam does not pass through the thicker disk material between foils. The position of the calibration foil is particularly important, since images of the calibration foil, taken periodically with back-lighting, provide the position of the nominal beam line on the camera pixel matrix. After final installation and alignment of the arm and disk, the calibration foil was rotated into the beam position and surveyed with a theodolite. The central calibration hole position with respect to the nominal beam line along the horn axis was measured with 0.3~mm precision.

The disk-rotation motor system is positioned 1.5~m underneath the helium vessel lid, above the iron and concrete shielding shown in Fig.~\ref{fig:otrpath}.  A long rigid steel shaft 
couples to a flexible steel shaft that follows a 90 degree bend to the arm supporting the foil disk.  A spline coupling is made at the end of the arm to another flexible shaft which runs along the arm and connects to the disk. The motor is connected to the shaft through a 100:1 gearbox, so that the disk rotates slowly. Access for maintenance or replacement of the motor system is possible through the window in the lid. 
 
The foil position can be determined in principle by counting the number of motor steps using its encoder. However, due to backlash in the flexible shaft system, alternate methods of ensuring precise foil positioning are necessary.  
The primary method to change foil positions is to run the motor until it is turned off by a micro-switch engaged by a machined titanium button on the disk. This type of micro-switch (model 6302-16 from Haydon), designed for extreme conditions, is used in high radiation environments at TRIUMF. 
 The switch position was adjusted so that it engages the button just before a foil is in the correct position. A steel ball-bearing plunger mechanism, spring-loaded against the surface of the disk, then falls into a matching machined depression in the disk, which locks it firmly into the correct position. Each foil position has a corresponding machined depression. Most of these features can be seen in Fig.~\ref{fig:foil_disk_rear}. When the motor is turned on again to move to the next foil position, the torque of the motor and shaft system is sufficient to start rotating the disk and bring the ball bearing out of the depression against the spring force. The foil position repeats with this method to 0.1~mm precision.

A backup system uses a pressurized helium gas line, which ends in a brass tube with an end face parallel to the back of the disk about 0.1~mm from the surface (see Fig.~\ref{fig:foil_disk_rear}). 
As the disk rotates the pressure is maintained until the correct foil position is reached, at which point the tube end encounters a hole through the disk, reducing the pressure and causing a pressure switch located outside the shielding to turn off the motor. 
This method is precise only to $\sim$2 mm, but sufficient to position a foil in the beam in the case of failure of both the  micro-switch and plunger. 
This same helium line can be used to remove any accumulated dust from the foil, since the gas passing through the hole in the disk is guided by a custom nozzle (see Fig.~\ref{fig:foil_disk}) to blow across the foil surface.

\begin{figure}
\begin{center}
\includegraphics[width=0.42\textwidth]{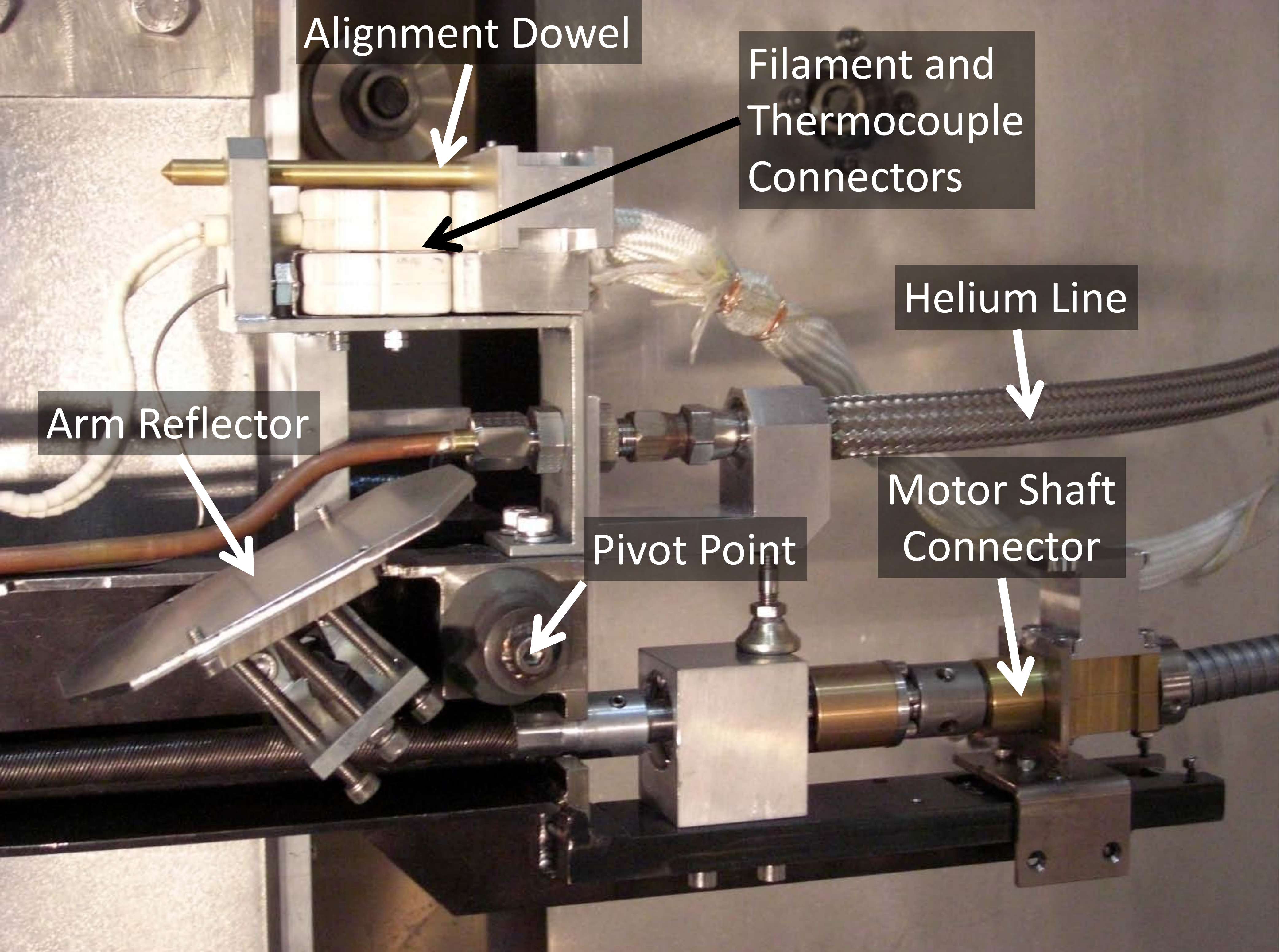}
\end{center}
\caption{The outside end of the arm.}
\label{fig:arm-end}
\end{figure}

In the event that a horn or the target require maintenance, the disk and/or arm can be removed or replaced by remote manipulators.
To remove the arm, the spline coupling of the motor shaft, the helium line and the ceramic connectors for the micro-switch and thermocouple must first be disconnected using the manipulators. A picture of the end of the arm is shown in Fig.~\ref{fig:arm-end}. 
For the T2K target replacement the arm can be rotated down by about $45^{\circ}$ on a pivot bearing on the outside leg without disconnecting the electrical, helium and spline connections. Using a mechanism on the inner leg the arm can be locked back into the correct horizontal position.
 
\subsection{The Mechanical Design of the Optical System}

\begin{figure}
\begin{center}
\includegraphics[width=0.42\textwidth]{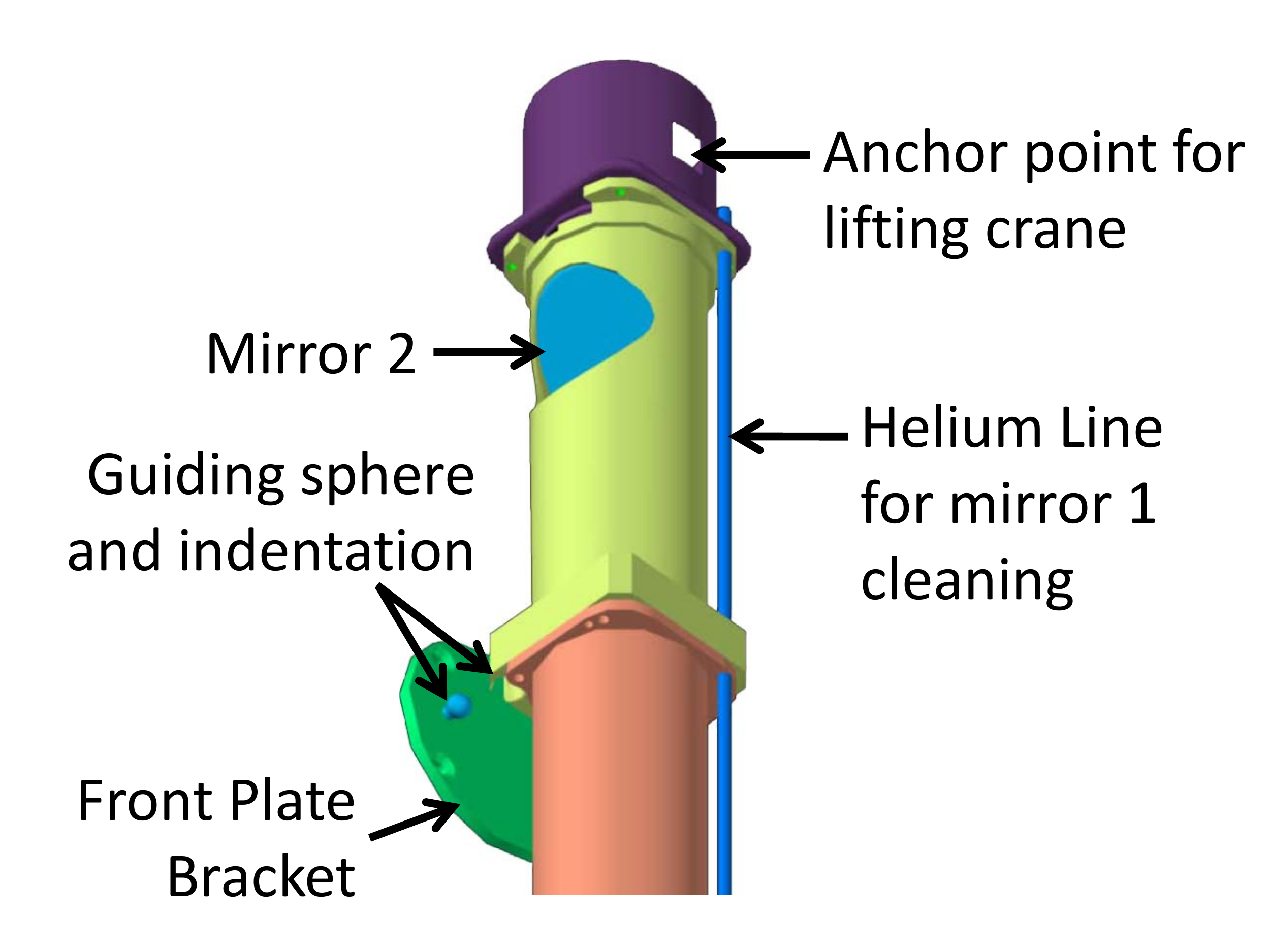}
\end{center}
\caption{Mirror tube system.}
\label{fig:mirror-tube}
\end{figure}

Mirrors 1 and 2 along the OTR light path (see Fig.~\ref{fig:otrpath} and also Fig.~\ref{fig:OTR_downstream_schematic}) are mounted at either end of a long steel tube and mirror~3 is at the bottom end of a second shorter steel tube. Some of the mirror tube details are shown schematically in Fig.~\ref{fig:mirror-tube}.
Each mirror can be rotationally adjusted about two axes. The tubes are mounted precisely on ball mounts protruding from flanges attached to the back of the front plate of the support module. They can be lifted out and replaced by crane from the lid of the helium vessel through ports directly above the tubes. 

\subsection{The Calibration Lighting Systems}

\begin{figure}[t]
\begin{center}
\includegraphics[width=0.42\textwidth]{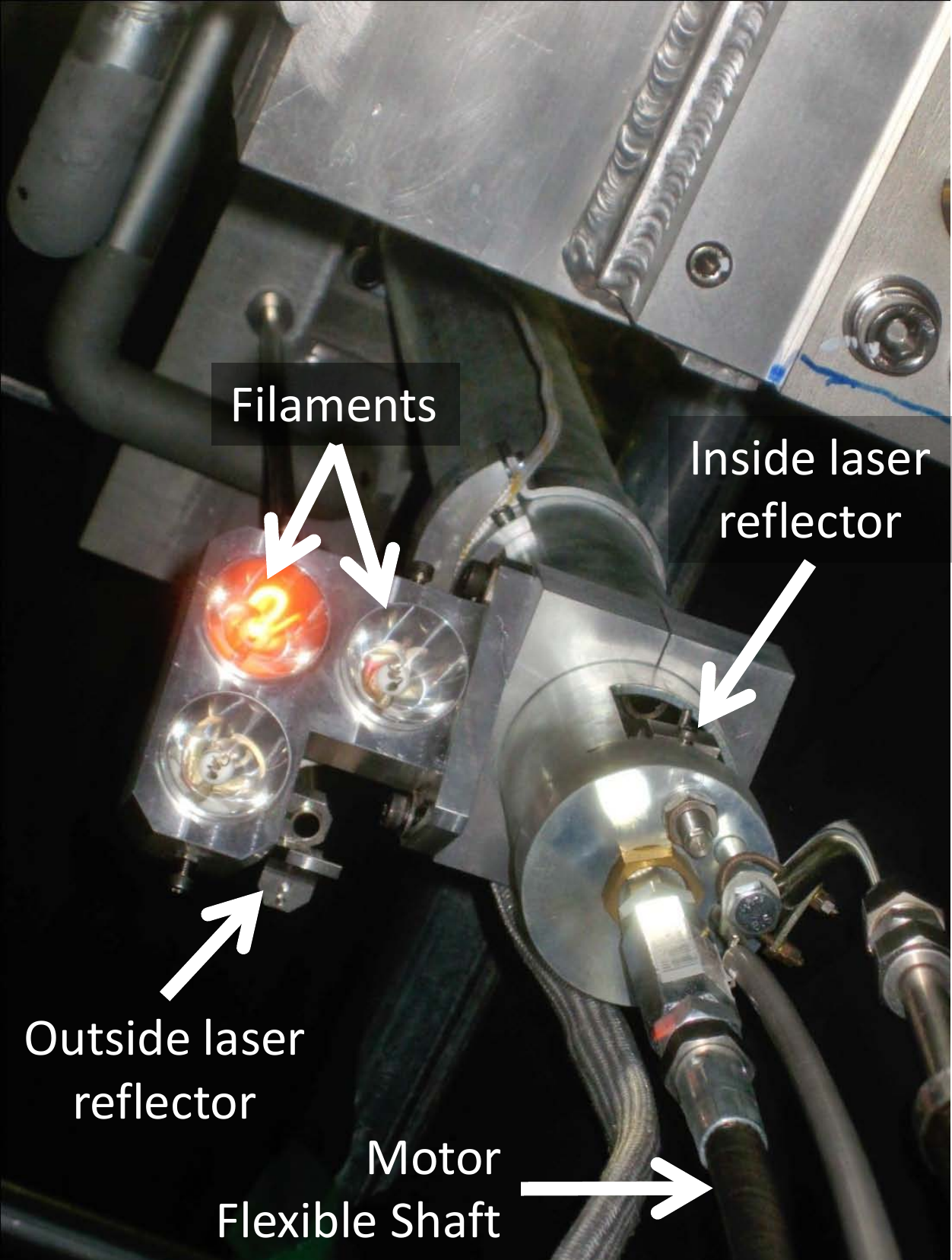}
\end{center}
\caption{A picture from below showing the filament lights, laser reflectors and coupling of the vertical to flexible shafts of the disk rotation system.}
\label{fig:filament-lamps}
\end{figure}

There are three lighting systems used for calibration by lighting the calibration foil from behind. 
Two of them use small red LED lasers (Sanyo DL3147-060,  650~nm, 7~mW), one mounted just above the fused silica window (outside laser) in the lid and the other mounted near the motor which drives the disk rotation (inside laser). 
There are two internally electro-polished steel tubes (10 and 13~mm diameter) which guide the laser light down to the region at the bottom of the tube extending from the front plate of the horn support module, seen on the right side of Fig.~\ref{fig:OTR_upstream_schematic}. 
Two small steel reflectors guide the laser light to another reflector on the arm, which can be seen in Fig.~\ref{fig:arm-end}. 
The third system is a set of filament lights (3 for redundancy), custom built with Alchrome wire coils in parabolic reflectors. They are installed in the same region as the laser reflectors and also point at the arm reflector.
Fig.~\ref{fig:filament-lamps} shows a picture of this region. The filament coil is operated at 12~A current, causing the wire coil to glow with sufficient light output.

The filament lights are not accessible after the beam has been running, and they cannot be replaced by the remote manipulators. The outside laser can be replaced whenever the beam is off for a brief period, but the inside laser near the motor can only be replaced in a long shutdown when the helium vessel lid has been removed.

\subsection{Alignment}
\begin{figure*}
\begin{center}
\subfigure[This diagram shows the aligment procedure for the mirrors.]{
\includegraphics[width = 0.42\textwidth]{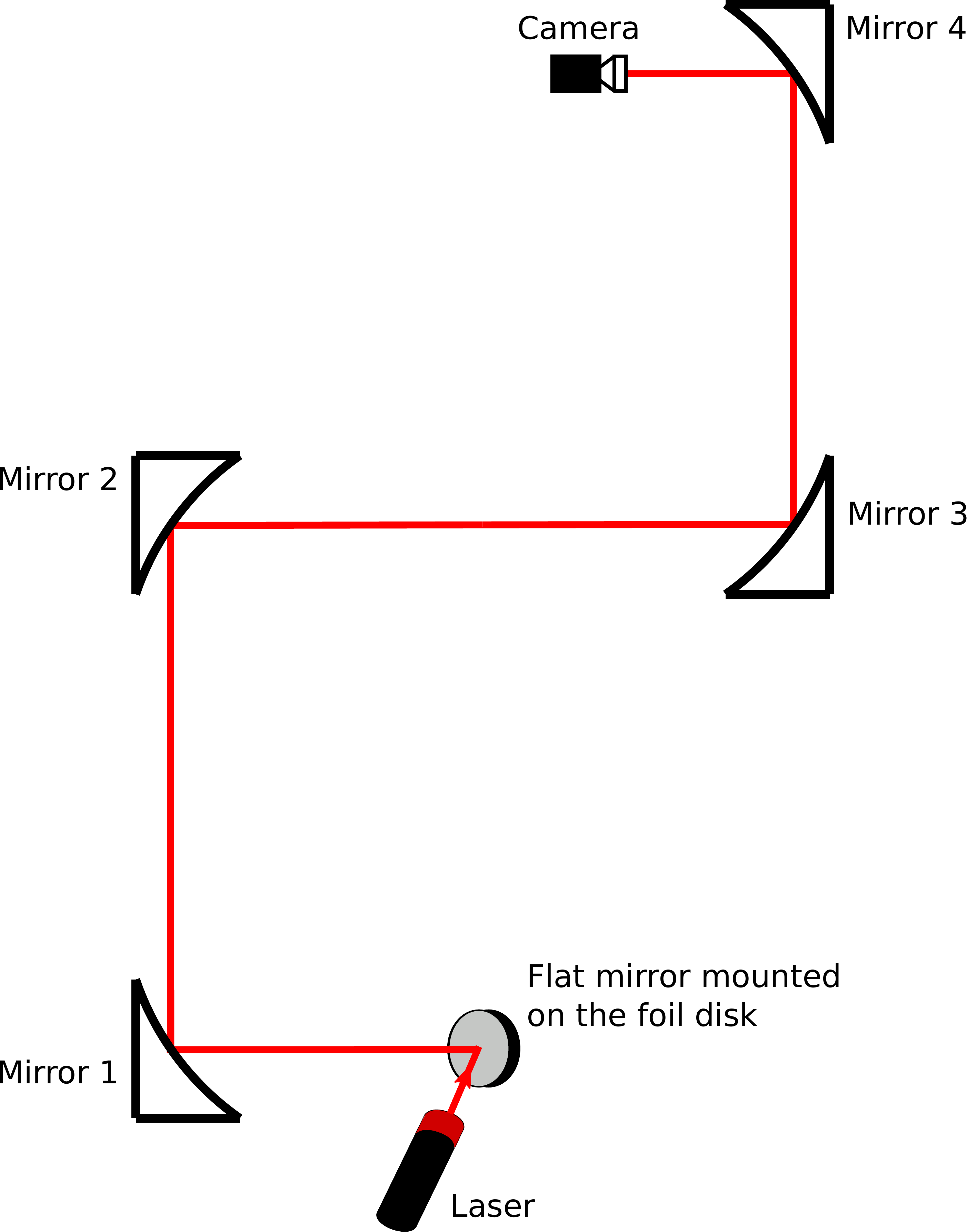}
\label{fig:OTR_align1}}
\subfigure[This diagram shows the aligment procedure for the arm reflector.]{
\includegraphics[width = 0.42\textwidth]{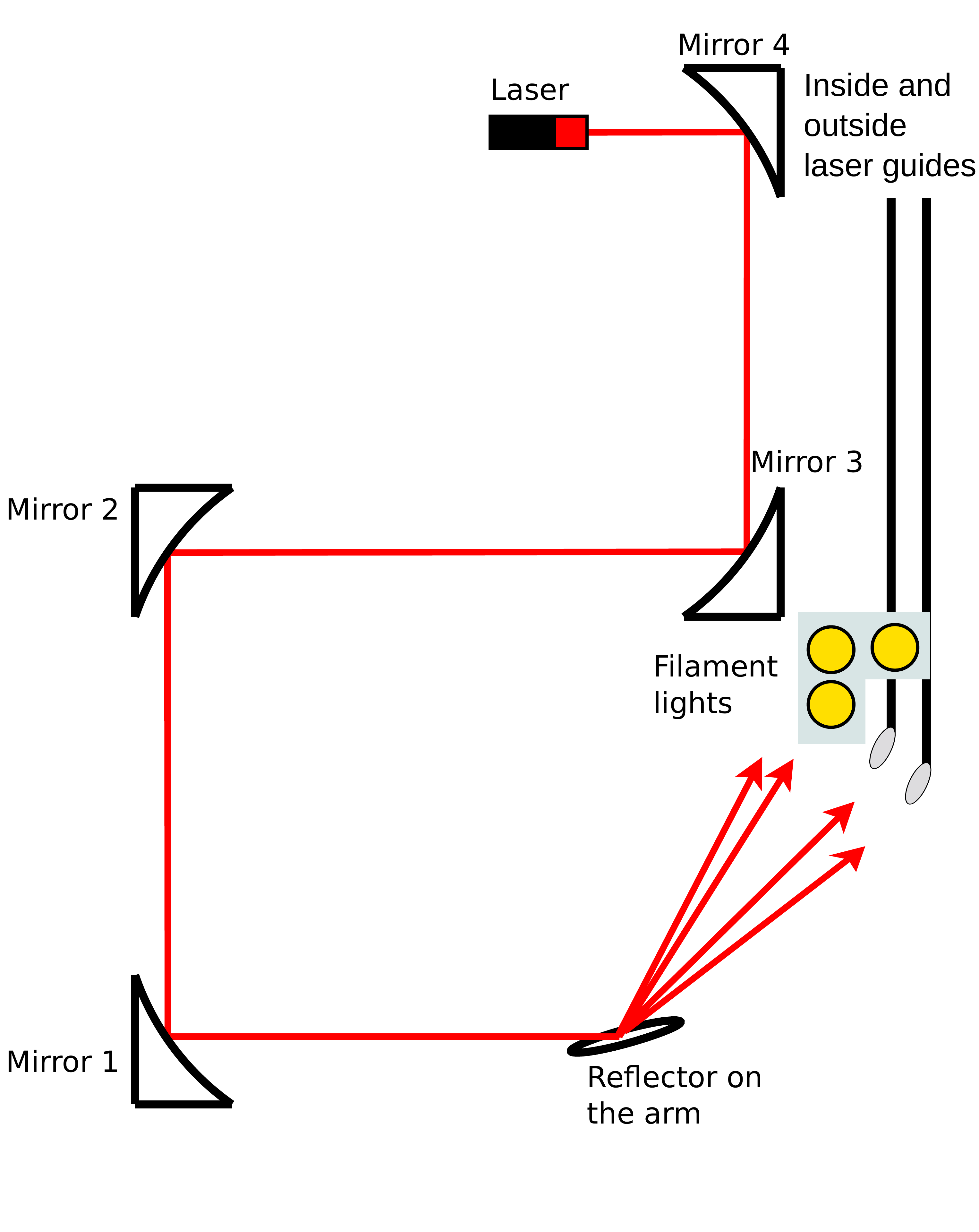}
\label{fig:OTR_align2}}
\end{center}
\caption{Figures illustrating the aligment of the OTR system components.}
\label{fig:otr_align}
\end{figure*}

The optical system was aligned using a laser. A special mount was made to attach the laser in front of the empty foil position in the disk. The laser was aligned to point along the beam line through the horn. A plane mirror was then attached parallel to the disk to reflect the laser light at 90~degrees along the OTR light path (see Fig.~\ref{fig:OTR_align1}). The mirrors were then adjusted one by one so that the laser light followed the correct central path through the optical system to the camera.  This was greatly aided by small marks made the center of each parabolic mirror during their manufacture.  The arm reflector was aligned by shining a laser back from the camera position through the optical system, through the empty foil position onto the reflector and up to the filament lamp region (see Fig.~\ref{fig:OTR_align2}). 

\subsection{Spare OTR System}

In order to be prepared for possible future problems with the target region of T2K, a spare horn, support module, target and OTR system have been built. The spare OTR system has been installed, aligned, and calibrated on the spare horn and support module.

\section{Data Acquisition and Slow Control}
\label{sec:daq}

\subsection{OTR DAQ system}
The main functions of the data acquisition system (DAQ) are to trigger, collect and process the image data. 

\begin{figure*}
\begin{center}
\includegraphics[width = 0.8\textwidth]{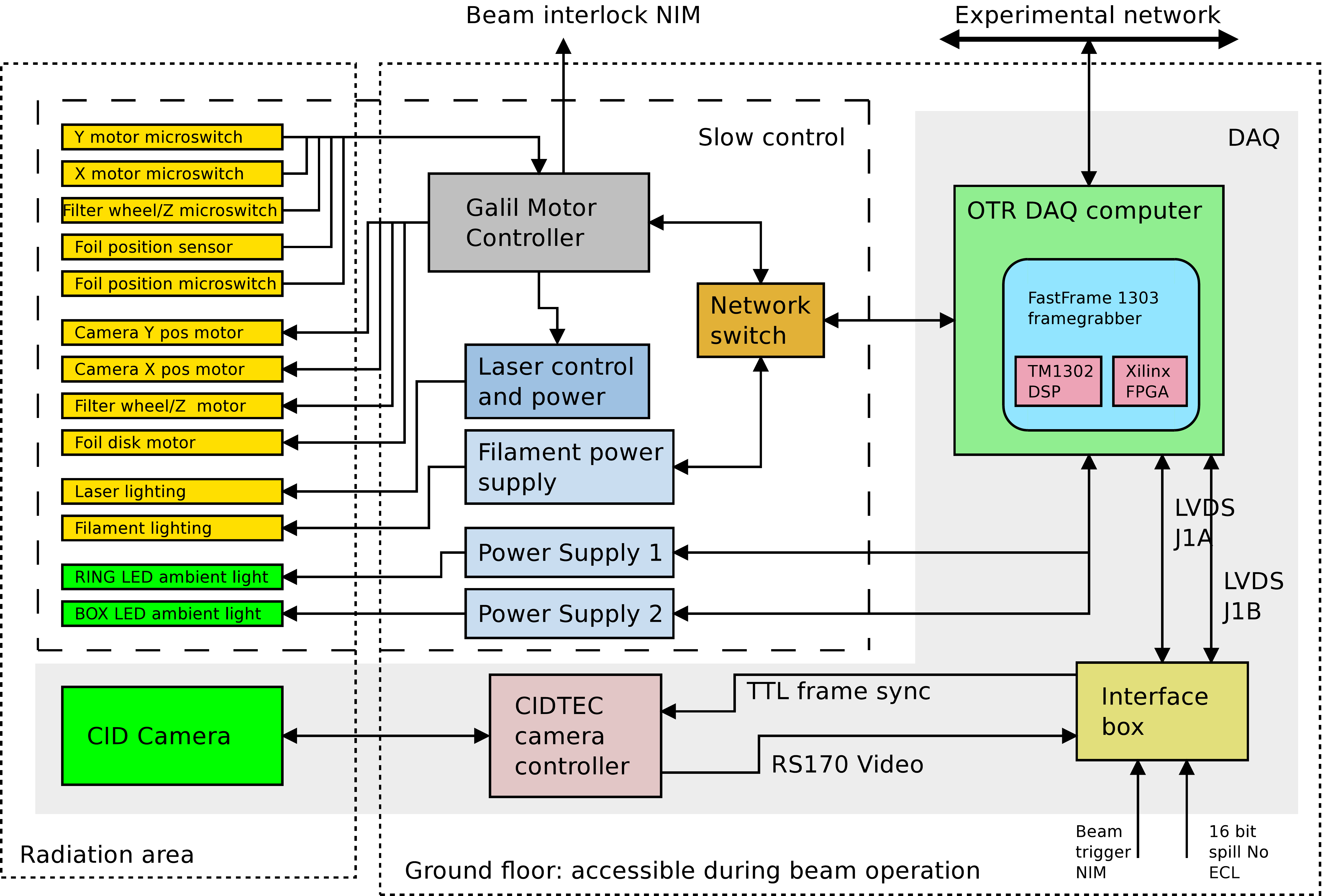}
\caption{OTR monitor slow control and DAQ system.}
\label{fig:otrdaqdia}
\end{center}
\end{figure*}

The components of the DAQ system are illustrated in Fig.~\ref{fig:otrdaqdia}. The trigger signals and the image acquisition controls are handled by a FPGA chip located on a frame-grabber board which interfaces to a host DAQ computer via a PCI bus. Also located on the board is a TriMedia TM1302 digital signal processor (DSP) which is responsible for the transfer of the digitized image frames to the host computer. Configuration of the FPGA registers is done via TriMedia software. 

\begin{figure}[ht]
\begin{center}
\includegraphics[width = 0.47\textwidth]{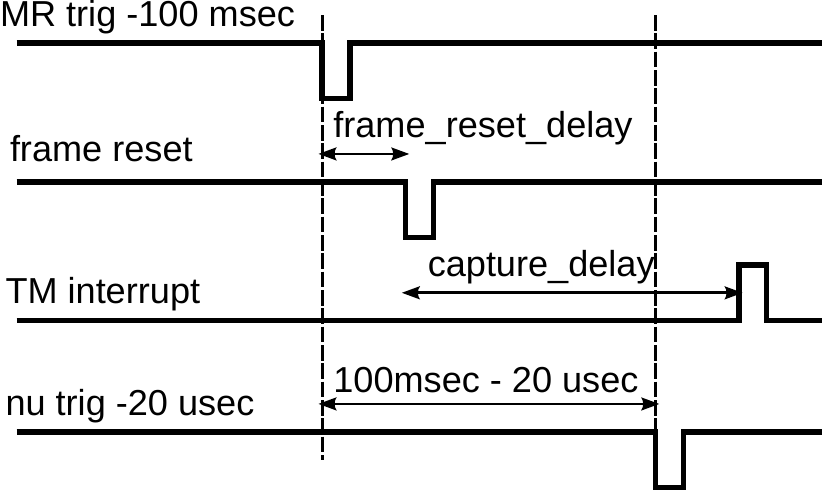}
\caption{Image acquisition signals.}
\label{fig:otrtimechart}
\end{center}
\end{figure}

The analog video signal from the camera is acquired and digitized by the frame-grabber. Fig.~\ref{fig:otrtimechart} shows timing signals important for the DAQ operation. The readout cycle is initiated with a pre-trigger which arrives 100 msec before a possible proton beam extraction to the neutrino beam-line. Following its arrival after a programmable delay (\verb+frame_reset_delay+) a frame reset signal is issued to the camera which synchronizes it and the readout circuitry to the expected spill arrival time. The trigger signaling the beam extraction to the neutrino beam-line arrives 20 $\mu$sec before the spill. After it is received and while the camera data are acquired and digitized by the frame-grabber an interrupt signal (TM interrupt) is sent to the DSP. The timing of the interrupt relative to the frame reset is configurable (\verb+capture_delay+). The interrupt signal informs the DSP that the next available frame will contain the spill data and should be moved from the internal memory buffers to a dedicated memory address on the host computer. After moving the spill image, the DSP also transfers the image data from the two subsequent frames. These images are later used for the pedestal subtraction (see Section~\ref{sec:imgana}).

Once the images are copied to the memory of the host computer, a MIDAS-based \cite{midas}, front-end application compresses and sends the image data to a dedicated event server that handles the distribution of the monitor data for online analysis and archiving. 
\begin{figure*}[p]
\begin{center}
\includegraphics[width = 0.85\textwidth]{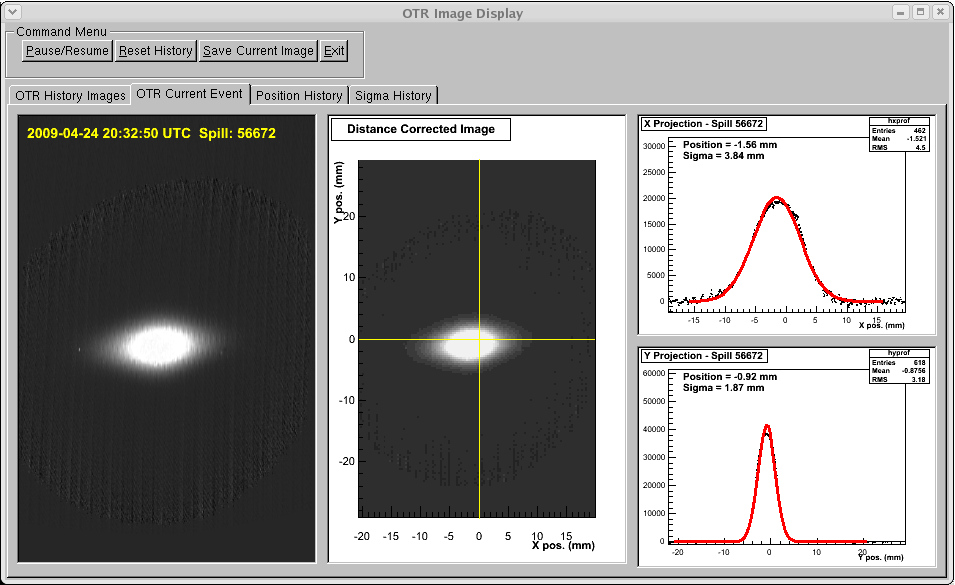}
\caption{OTR online event display showing the proton beam on target. The center of the target is marked by a yellow cross-hair in the middle panel.}
\label{fig:onlinedisplay}
\end{center}
\end{figure*}

The OTR online monitoring program displaying an event from one of the first proton beam extractions to the neutrino beam-line at J-PARC is shown in Fig.~\ref{fig:onlinedisplay}. The center of the T2K target is marked by a yellow cross-hair in the middle panel. The two panels on the right show in the horizontal and vertical projections the results of a 2D fit that extracts the beam position and width. These values along with the measurements from the upstream proton beam monitors are used to determine whether the beam position and width are within tolerance. If they are outside tolerance, an abort signal is sent to prevent further extraction from the main accelerator ring to the neutrino beam-line.

\subsection{Slow Control}
The OTR monitor has a number of remotely controlled motors to move different components of the detector. In addition, the lighting system for the periodic calibration of the optics requires a number of power supplies. These elements are part of the slow control system (Fig.~\ref{fig:otrdaqdia}) which is managed using MIDAS.

The hardware control of the disk, filter wheel and camera stage motors is done using a Galil motor control module. The unit also collects the status from the pressure sensor and the micro-switches on the motors. The state of the disk micro-switch is constantly monitored. To avoid possible proton beam extraction during the foil disk rotation or when the disk is not properly positioned, a beam interlock signal is generated when it is not engaged.

\section{Image Correction and Analysis}
\label{sec:icalib} 

\subsection{Efficiency Correction}
\label{subsec:efficiency}

\begin{figure*}[h]
\begin{center}
\subfigure[Simulated efficiency of the optical system.]{
\includegraphics[width = 0.45\textwidth]{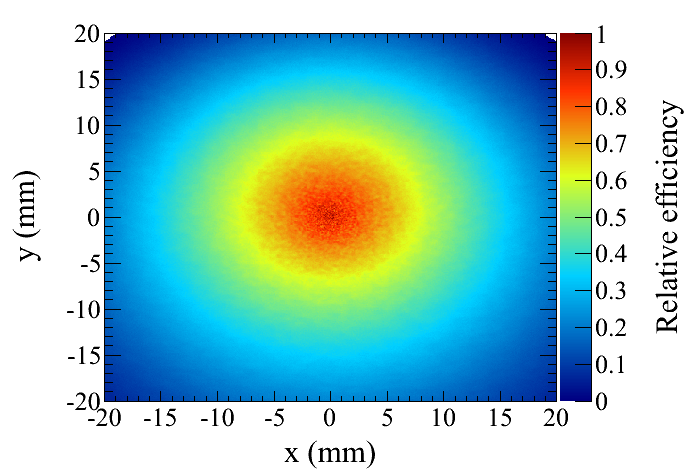}
\label{fig:effsim}}
\subfigure[Efficiency measured with the integrating sphere.]{
\includegraphics[width = 0.45\textwidth]{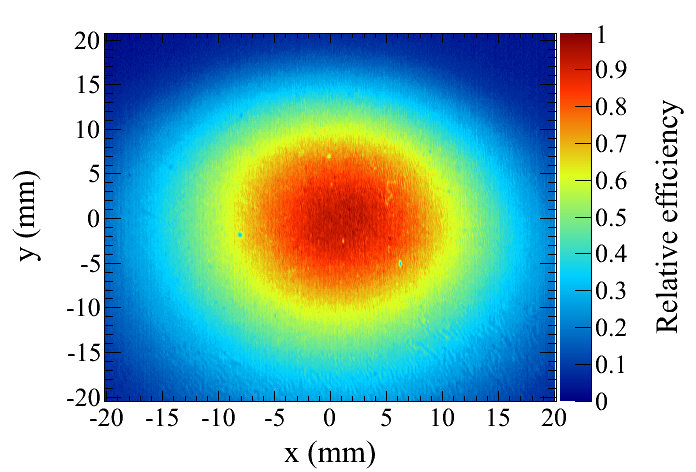}
\label{fig:effimg}}
\caption{Simulated (a) and measured (b) light collection efficiencies. Each distribution is normalized by the maximum effeciency which is at the center of the image.}
\label{fig:eff}
\end{center}
\end{figure*}

Ray tracing simulations of the optical system revealed that within $\pm$15 mm of the foil center, the relative light collection efficiency varies by more than 50\% (Fig.~\ref{fig:effsim}). 
This introduces a bias into the reconstructed beam position and width, since the
variation of the light collection efficiency is significant over the size of the
beam spot. 

\begin{figure}[t]
\begin{center}
\includegraphics[width = 0.45\textwidth]{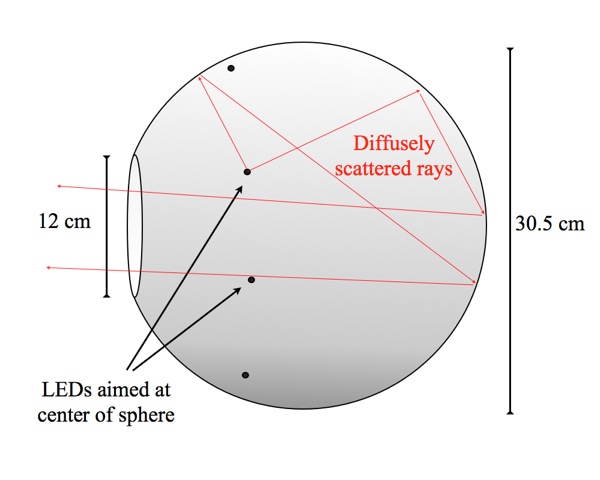}
\end{center}
\caption{A schematic of the integrating sphere used to measure the light collection efficiency.}
\label{fig:integrating_sphere}
\end{figure}

An integrating sphere  was used to provide a uniform light source in order to measure the light collection efficiency of the system. This hollow cardboard sphere, shown in Fig.~\ref{fig:integrating_sphere}, is 30.5~cm in diameter and painted white on the inside. A 12~cm opening at one pole of the sphere is lit internally by an off-equatorial ring of 8 laser diodes.  There is no direct path between the diodes and the opening, so all of the light exiting the sphere results from diffuse reflections off  the rough inner surface of the sphere.
This inexpensive device gives excellent performance, with light output measured to be uniform in intensity 
within 5\% across the entire opening. Prior to the installation of the horn module into the helium vessel, the sphere 
was positioned at the foil location with the opening facing mirror~1.
Images of the light through the optical system were taken with the camera system, as shown in Fig.~\ref{fig:effimg}, and are used to correct for efficiency.

During periodic calibration runs, images are collected with the back-lighting from the filament and laser light sources illuminating the empty foil slot. 
The back-lighting is less uniform than the light from the globe,  but these images are used to monitor the stability of the light collection efficiency over time.

\subsection{Distortion Correction}
\label{subsec:distortion}
\begin{figure}[htb]
\begin{center}
\includegraphics[width=0.45\textwidth]{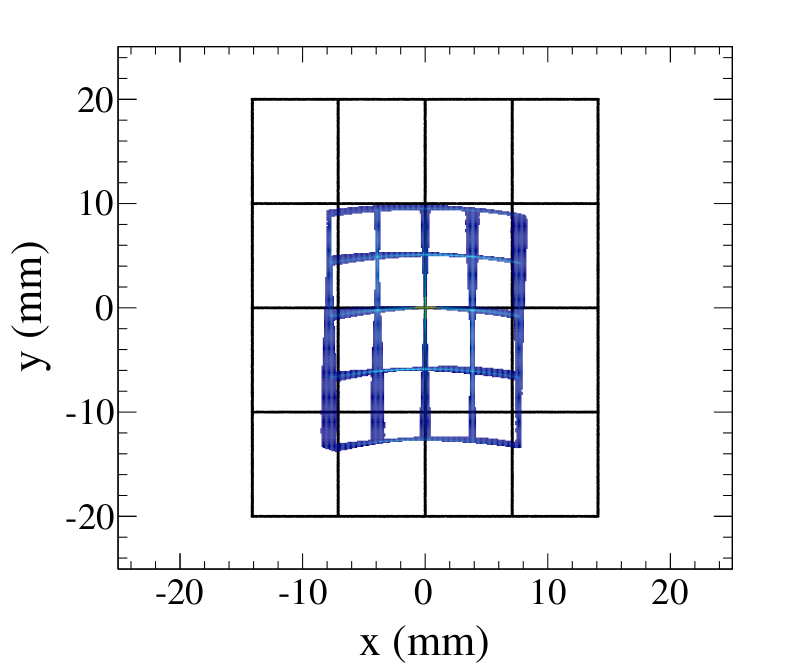}
\end{center}
\caption{The thin black lines show a grid of point sources placed on the foil surface, and the wider lines show the resulting image after simulated rays were traced through the optical system.  The size of the image is also reduced by 54\% compared to the original due to the smaller focal length of mirror 4 compared to the first 3 mirrors.}
\label{fig:distortion55}
\end{figure}

The optical system introduces a certain amount of distortion to the image that can bias measurements and therefore requires correction.  Fig.~\ref{fig:distortion55} shows what happens to a square grid of points on the surface of the foil after the rays are traced through the optical system. The distortion is minimal at the centre of the image, but increases near the edges, and is more pronounced in the vertical direction due to the nature of the parabolic mirrors.

\begin{figure}[htb]
\begin{center}
\includegraphics[width=0.4\textwidth]{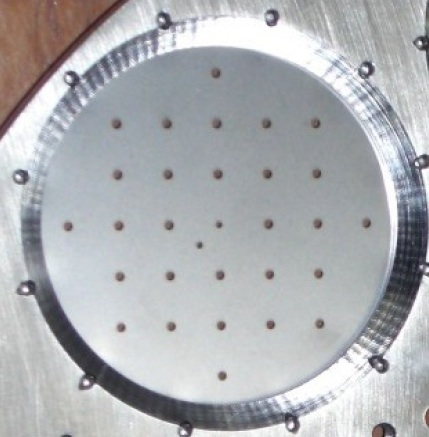}
\end{center}
\caption{The calibration foil with a grid of laser-machined holes.}
\label{fig:foilgrid}
\end{figure}

The calibration foil, shown in Fig.~\ref{fig:foilgrid}, has laser-machined holes arranged in a grid with 7~mm spacing. The holes are 1.2~mm in diameter except for the 2 central holes, which are 0.8~mm in diameter. The relative locations of holes are known to a precision better than 0.1~mm and the central hole position is known with respect to the beam line axis to 0.3 mm, as discussed in Section~\ref{sec:mechanical}.   This hole provides an absolute position reference point for the images recorded by the camera. 

\begin{figure*}[ht]
\centering
\subfigure[Image of the back-lit calibration foil showing the characteristic distortion introduced by the optics.]{
\includegraphics[width = 0.3\textwidth]{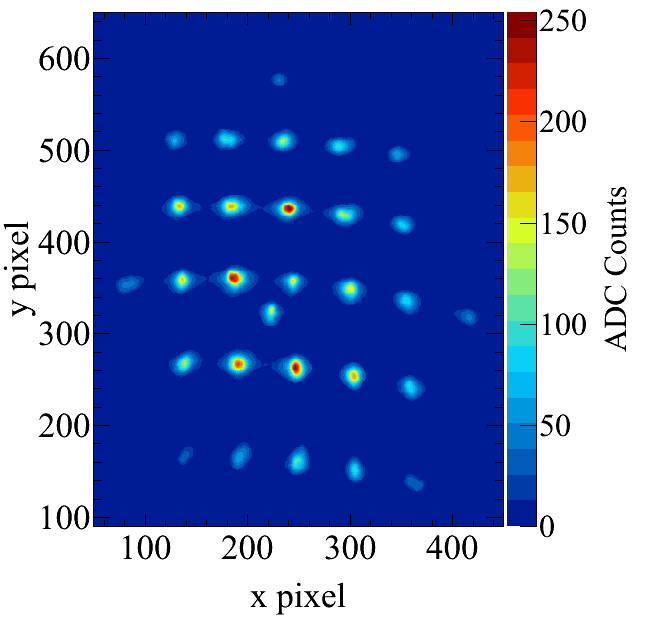}
\label{fig:otrdistimg_1}}
\subfigure[Image of the distorted calibration foil pattern with centroids of the holes marked by black triangles.]{
\includegraphics[width = 0.3\textwidth]{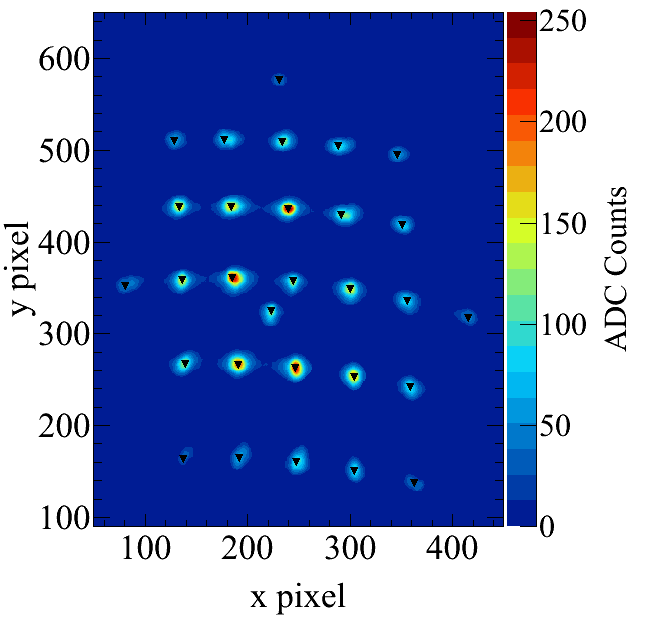}
\label{fig:otrdistimg_2}}
\subfigure[Image of the back-lit calibration foil after the distortion correction with the true hole pattern supperimposed.]{
\includegraphics[width = 0.3\textwidth]{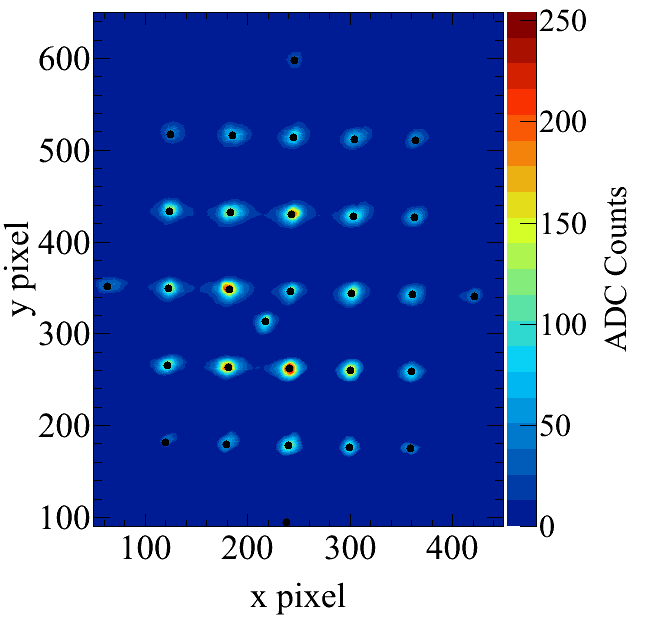}
\label{fig:otrdistimg_3}}
\caption{Illustration of distortion correction for the calibration foil image.}
\label{fig:otrdistimg}
\end{figure*}

The calibration lighting systems are used to back-light the calibration foil and take images of its hole pattern. In addition to referencing the position of the beam axis, the grid of holes is used to correct for distortion in the optical system.  The calibration images are taken regularly during periods of no beam operation so that any movement of the hole pattern with respect to the camera pixels due to changes in the optical path can be followed and corrected.  
Small variations with time due to such effects as temperature changes are expected. The optical system has shown remarkably stability at the level of 0.1~mm to date. 

The image of the pattern on the back-lit calibration foil is sensitive to the angles of the incident light. This results in $~0.2$ to $~0.25$ mm shifts in the imaged hole position between different light sources. This was estimated from an unbiased image of the calibration pattern, obtained by placing a light source directly behind the calibration foil on the optical axis (the axis being defined by the line of sight between the central hole of the foil and the center of mirror 1).

The distortion of the imaged calibration foil pattern is visible in Fig.~\ref{fig:otrdistimg_1} and follows the expectations from simulations (cf. Fig.~\ref{fig:distortion55}). These calibration images are analyzed to obtain the centroids of the holes marked by black triangles in Fig.~\ref{fig:otrdistimg_2}. This information in comparison to the true hole positions is used to build a transformation map to correct image distortions. As an example, Fig.~\ref{fig:otrdistimg_3} shows the image of the calibration foil pattern after the distortion correction with superimposed true positions of the holes. 

\subsection{Image Analysis}
\label{sec:imgana}
The position and width of the proton beam as well as the total light yield from a given foil is determined spill by spill. The  analysis begins by subtracting the pedestal from each spill image. The pedestal values are obtained from the two images taken immediately after the spill image. Following the pedestal subtraction a number of corrections are performed:
\begin{list}{\labelitemi}{\itemsep=0pt\parsep=0pt}
\renewcommand{\labelitemi}{{\tiny$\bullet$}}
\item{correction due to the non-uniform light collection efficiency of the optical system using the map obtained with the integrating sphere}
\item{correction due to the charge decay in the camera sensor discussed in Section~\ref{subsec:camera}}
\item{correction of the image distortions with the transformation map obtained from the back-lit image of the calibration foil.}
\end{list}
After the image corrections, a two-dimensional fit to the data is performed to extract the proton beam position and profile, and the total light yield.

\section{Performance}
\label{sec:perform}
The OTR monitor has been in operation with proton beam intensities ranging from 
$1\times10^{11}$ to $1\times10^{14}$ protons per spill.
As has been mentioned, for very low intensity operation during the commissioning of the beam-line the 
fluorescent light from the ceramic wafer was used.
For intensities above $1\times10^{12}$ the OTR light
production was large enough to image the proton beam and OTR light 
has been imaged for both titanium and aluminum alloy target foils.

\subsection{Ceramic Wafer Performance}
\label{subsec:waferperform}
\begin{figure}
\begin{center}
\includegraphics[width=60mm]{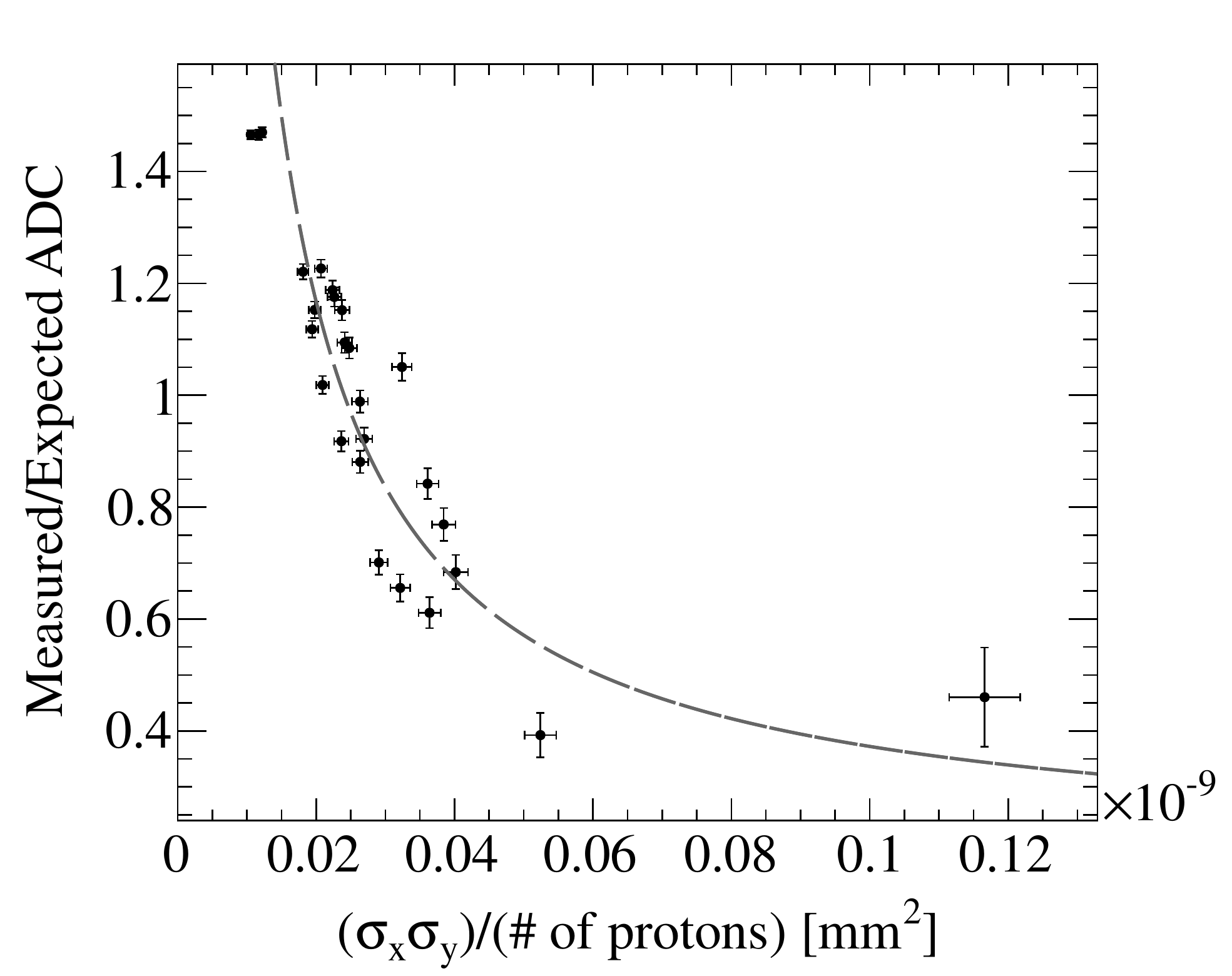}
\end{center}
\caption{Ratio of measured over expected light yield from the fluorescent disk  as a function
of $\sigma_x \sigma_y/$(number of protons) where the expected light yield assumes a linear
response for the fluorescence.  The line is the fitted expectation for light production that
is quadratic with the number of protons.}
\label{fluor_nonlin}
\end{figure}

The expected light production from the ceramic wafer and camera response
is estimated in a similar fashion to that for OTR light.  
In the case of the fluorescent light, the production is isotropic and 
the spectrum is sharply peaked with lines at 692.9~nm and 694.3~nm and the photon yield per deposited energy has been measured for
1~MeV stopping H$^{+}$~\cite{fluorescence} as:
\begin{linenomath*}
\begin{equation}
E_{dep} = 4.94\times10^{4} \frac{\mbox{photons}}{\mbox{MeV}}.
\end{equation}
\end{linenomath*}
The expectation of $6.0\times10^{-2}$ electrons per proton is larger than the measured yield of $0.9\times10^{-2}$ electrons per proton 
measured with $3.6\times10^{11}$ protons per spill. The discrepancy may arise from different light production for 30 GeV protons, unaccounted imperfections in the optical system or non-linear light production.  As illustrated in
Fig.~\ref{fluor_nonlin}, the response normalized by the expectation based
on the number of protons increases as the beam becomes narrower, or the intensity
 increases.  The nonlinear production of fluorescent light 
makes it unsuitable for beam width measurements without a correction for the 
non-linearity.

\subsection{Titanium and Aluminum Foil Performance}

\begin{figure*}[htb]
\begin{center}
\includegraphics[width=0.43\textwidth]{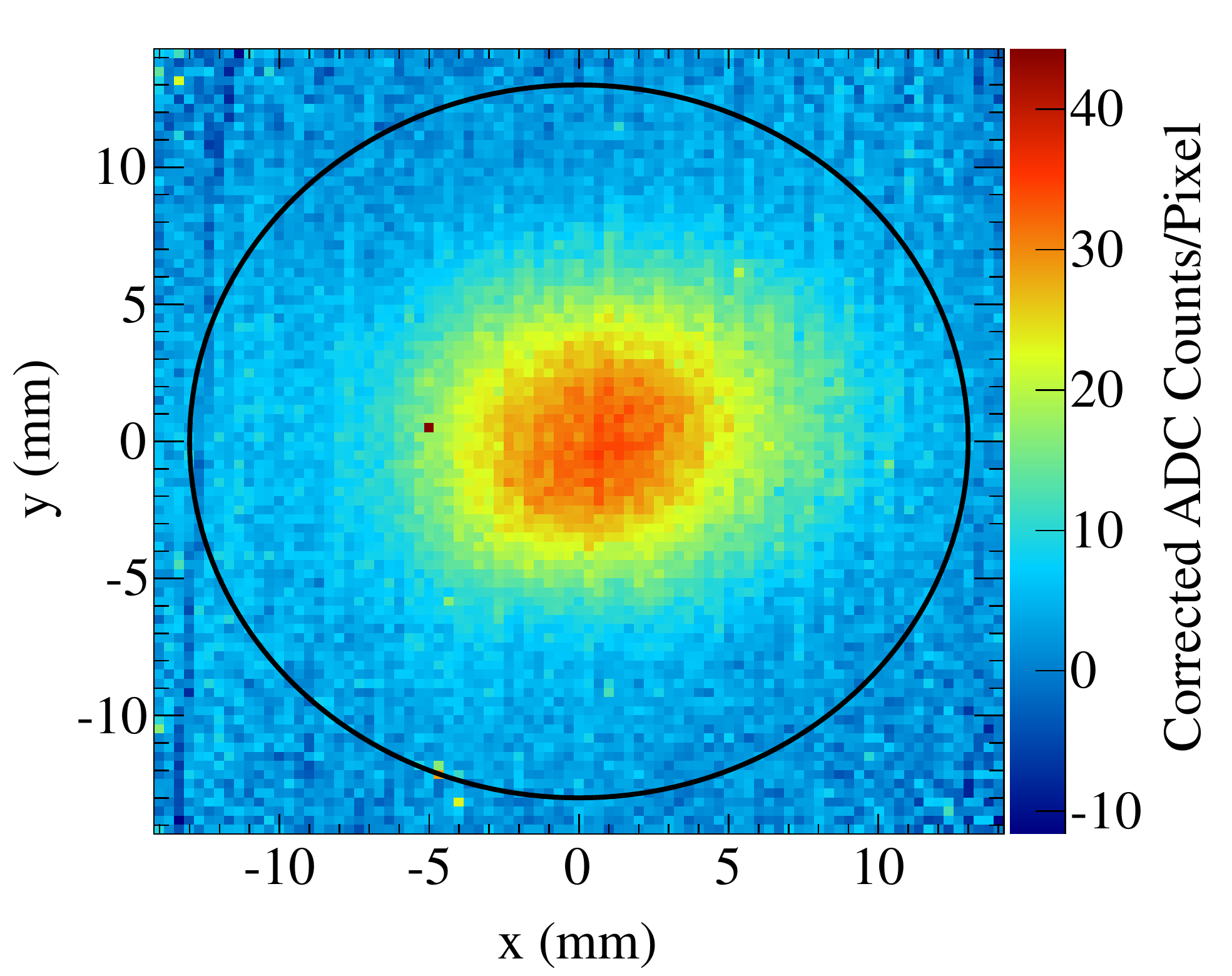}
\includegraphics[width=0.43\textwidth]{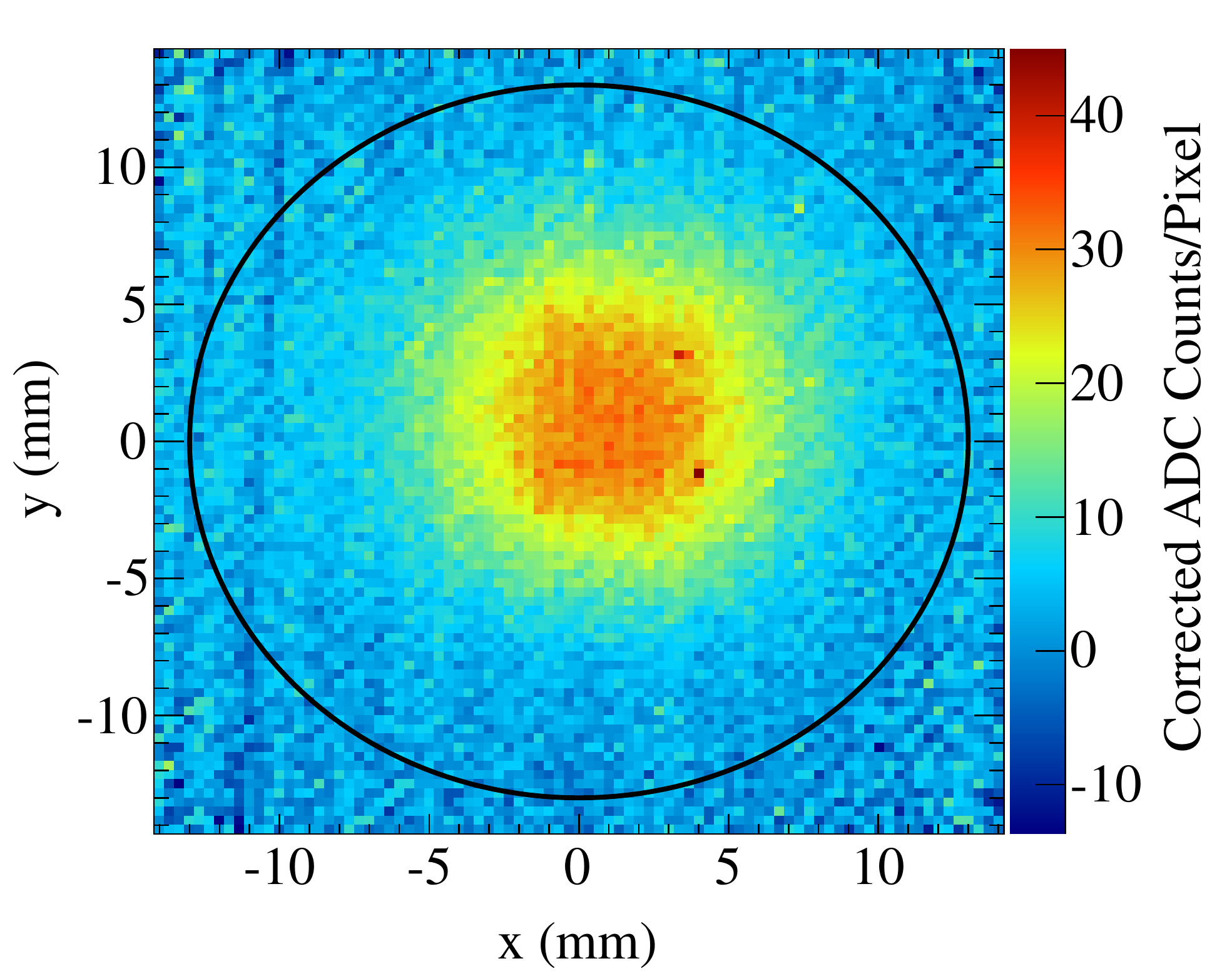}
\end{center}
\caption{Example OTR images of the beam with nominal beam conditions using
the titanium (left) and aluminum (right) alloy target foils.  The images are
taken with $9.0\times10^{13}$ and $2.3\times10^{13}$ protons per spill, respectively.  The
black circle indicates the location of the target edge.}
\label{otr_spills}
\end{figure*}

Data collected during physics runs has been carried out by imaging OTR light
from the titanium and aluminum alloy foils.  Fig.~\ref{otr_spills}
shows example spills from data taking for both types of foil. The light yield from aluminum is found to be 2.8 times greater than that from
titanium.

\begin{figure}
\begin{center}
\includegraphics[width=60mm]{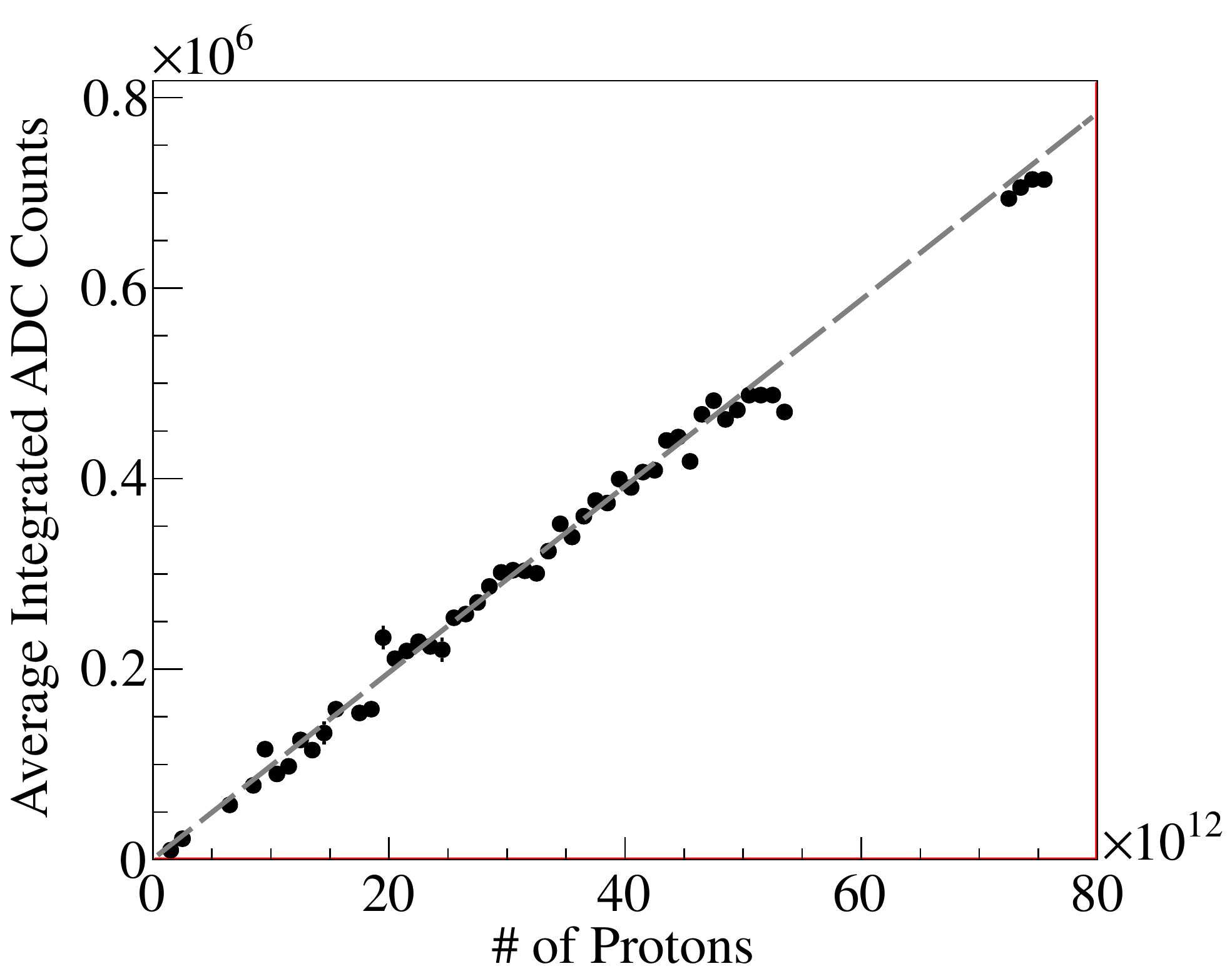}
\end{center}
\caption{Camera response as function of protons per beam spill for the Ti
alloy foil. The dashed line is the fit to the data.}
\label{otr_lin}
\end{figure}

The expected camera response using titanium  was 
calculated in Section~\ref{sec:photonyield} to be $2.1\times10^{-5}$ electrons per proton.  
The measured value of $4.8\times10^{-6}$ electrons per proton is 23\% of the expectation and in line with the deficit observed on the fluorescent foil, suggesting that the deficit is due to imperfections in the optical system such as the roughness of the foil and mirror surfaces, lower than expected mirror reflectivity or collection of dust on the mirrors.
The OTR light production as a function of proton beam intensity is measured to be linear over more than one 
order of magnitude as shown in Fig.~\ref{otr_lin}. 

An upper limit on the  position 
resolution is obtained by measuring the RMS of OTR measurements
for 50 consecutive spills taken with $7.4\times10^{13}$ protons
per spill.  Here it is assumed that beam fluctuations are negligible 
compared to the monitor resolution.  Measured resolutions are shown
in Table~\ref{otr_res} and are all $<100$ $\mu$m.  An improvement in 
resolution with increased beam intensity is observed as the signal to
electronic noise ratio improves.

\begin{table}
\begin{center}
\caption{Measurement resolutions of the OTR monitor using the titanium target with
$7.4\times10^{13}$ protons per spill}
\vspace{0.1in}
\begin{tabular}{cc}
\hline
Measurement & Resolution ($\mu$m) \\ \hline \hline
$x$ & 69 \\ 
$y$ & 85 \\ 
$\sigma_{x}$ & 68 \\ 
$\sigma_{y}$ & 54 \\ \hline
\end{tabular}
\label{otr_res}
\end{center}
\end{table}

\begin{figure}
\begin{center}
\includegraphics[width=80mm]{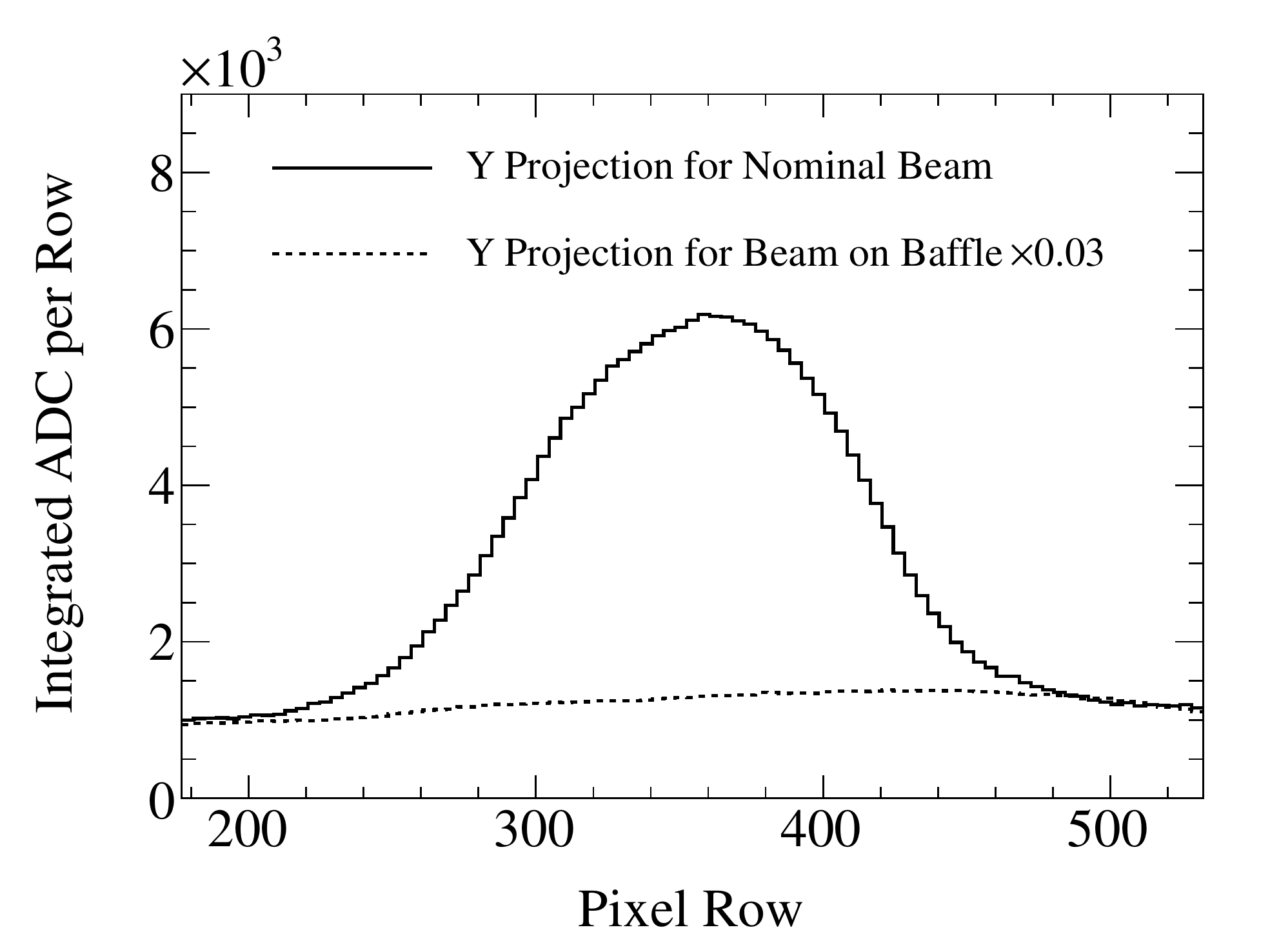}
\end{center}
\caption{The solid line shows the projection of OTR
light onto the y axis.   The dashed line shows the broad light level observed when the 
beam is directed onto the collimator, normalized to the background level
observed in OTR light images.}
\label{otr_bgnd_yproj}
\end{figure}

A broad source of background light
is observed in OTR images, as illustrated in Fig.~\ref{otr_bgnd_yproj}.  
Fig.~\ref{otr_bgnd_yproj} also shows the shape of the observed light when the proton beam is
directed onto the collimator located just upstream of the
OTR detector. This broad distribution, which is consistent with the tails of the light distribution seen when the beam is on target, may originate from secondary particles causing  scintillation of the helium gas or fluorescence of the ceramic materials near the target.

Measurements of the background light with optical filters selecting $\lambda>400$ nm 
and $\lambda<650$ nm are shown in Table~\ref{bgnd_filter_meas}.  The background light
has a broad spectrum with most light in the red or infrared.  This indicates that the scintillation light of helium is likely a small component of the background, since helium scintillation is strongly peaked at $<100$~nm~\cite{helium}.

\begin{table}
\begin{center}
\caption{The fraction of the background light in different
wavelength regions.}
\vspace{0.1in}
\begin{tabular}{cccc}
\hline
 & \multicolumn{3}{c}{Wavelength Range (nm)} \\ 
         &$<400$ & $400-650$ & $>650$  \\ \hline \hline
Fraction & 0.10 & 0.22 & 0.68 \\ \hline
\end{tabular}
\label{bgnd_filter_meas}
\end{center}
\end{table}

A test was made with a liquid crystal shutter that closed 300~$\mu$s after the beam pulse.  This had no significant effect on the amount of light collected, thus ruling out a large contribution to the background light from long-lived fluorescent sources.

\subsection{Systematic uncertainties}
The OTR monitor measures the beam properties with better than 100~$\mu$m resolution,
but systematic effects contribute larger errors to the total measurement 
uncertainties.  The largest source of uncertainty for the position measurement is
the alignment uncertainty from the survey of the calibration foil, 300~$\mu$m.  
Another significant source of uncertainty stems from the alignment of the light
sources used to back-light the calibration foil.  As discussed is Section~\ref{subsec:distortion}, since the light sources are not perfectly
aligned to the optical axis, the light at each calibration hole is not isotropic 
and a bias is introduced if the focusing is not perfect, introducing a shift of
of $200-250$ $\mu$m.   The entire magnitude of this shift is included as a systematic uncertainty.
\begin{table*}
\begin{center}
\caption{Sources of systematic uncertainties for single spill 
OTR measurements.}
\label{otr_syst_unc}
\vspace{0.1in}
\begin{tabular}{lcccc}
\hline
Source & $\delta x (\mu\mbox{m})$ & $\delta y (\mu\mbox{m})$ & $\delta \sigma_x (\mu\mbox{m})$ & $\delta \sigma_y (\mu\mbox{m})$ \\ \hline \hline
Calib. foil alignment  & 302 & 300 &  87 & 102 \\ 
Signal model           &   5 &   3 & 436 & 376 \\ 
Background model       &  90 & 115 &  10 &  31 \\ 
Fitter bias            &   4 &  15 & 105 & 140 \\ 
Calib. light alignment & 210 & 251 &  46 &  38 \\ 
Pixel charge decay     & 101 &  84 &  19 &  30 \\ 
Distortion correction  &  29 &  39 &  83 & 111 \\ 
Others sources         &  95 &  57 &  79 &  85 \\ \hline \hline
Total                  & 404 & 432 & 473 & 441 \\ \hline
\end{tabular}
\end{center}
\end{table*}

The beam width measurement uncertainty is dominated by the choice of signal function
used to fit the data.  Since a background source of light is present and modeled,
there is some freedom of choice for the signal model.  A two dimensional Gaussian 
model is used by default, but other models are consistent with the data.  The uncertainty
introduced by the choice of model is $370-440$ $\mu$m. 

Table~\ref{otr_syst_unc} lists the systematic uncertainties in the OTR measurements.
The total uncertainties are $<500$ $\mu$m for all measurements, matching the requirements
for the monitor.

\section{Radiation Damage Monitoring}

The radiation dose accumulated at the camera was measured by placing film badges near the 
camera during a run period when 9.59$\times10^{18}$ protons on target were accumulated. 
The measured dose of 1.34~Gy of gamma radiation extrapolates to a dose of 1.09~kGy for 
750~kW $\times$ 5$\times10^7$~sec of running, the T2K design exposure.  The dose from fast 
neutrons saturated the film badges at 25~mGy. The camera described in Section~\ref{subsec:camera} 
is rated to 10~kGy of total dose (gamma radiation) and it has since been replaced by a 
Thermo Fisher Scientific 8725DX model CID camera rated to 30~kGy of total dose (gamma radiation).

The direct measurement of the OTR light yield stability is precluded by variations in the 
camera gain.  Changes in the OTR light production due to beam related damage of the titanium foil are
evaluated by regularly comparing the light yield from the foil that is in the beam during data taking
to one of the spare titanium foils. Data taken during a period when $3.5\times10^{19}$ protons
on target were accumulated showed no significant change in the OTR light yield.

\section{Conclusions}
An Optical Transition Radiation monitor has been constructed for the proton beam-line of T2K experiment to operate in the high radiation environment near the T2K target. The monitor measures the position and profile to sub-millimeter precision. The optical transition radiation has been observed from aluminum and titanium alloy foils. The estimated resolution of the OTR monitor for the beam position and width variations is $\sim$100 $\mu$m while the absolute uncertainty on the these measurements is $\sim$ 500 $\mu$m. 

The beam position and angle at the T2K target are determined from the measurements of the OTR and the upstream position monitors described in \cite{t2k_nim}. The resultant overall uncertainty is $<1$ mm for the beam position and $<0.5$ mrad for beam angle at the target.

In conclusion, the OTR monitor has worked remarkably well in a challenging environment. It was critical for the tuning of the proton beam orbit during the start-up of the T2K experiment and provides valuable input for the physics analysis.

\section{Acknowledgments}
First and foremost, we would like to thank the T2K collaboration, the support staff at the J-PARC center, and the target station technicians. We would especially like to acknowledge the highly productive and enjoyable collaboration we had  with the T2K target station and beam group, including Martin Baldwin, Chris Densham, Michael Fitton, Atsuko Ichikawa, Hidekazu Kakuno, Takeshi Nakadaira, Ken Sakshita, Tetsuro Sekiguchi,  Masaru Tada, Mike Woodward, Yoshikazu Yamada, and Eric Zimmerman.   Special thanks to Matt Rooney who performed simulations to determine the thermal stresses in the foils.  We would like to acknowledge the contributions of Carl Ross and his group at NRC for providing electron beam time and invaluable assistance during the OTR prototype test, Brian Creber at B-Con engineering for helpful advice on the mirror production, the Institute for Optical Sciences at the University of Toronto, who kindly allowed us to use their laser machining facility, and  Breault Research Organization which allowed us to use their ASAP software in the early modeling of the optical system.  Thanks to undergraduate students Jordan Myslik, Eoin O'Dwyer, Stephen Ro, and Don Teo.  TRIUMF personnel helped with various aspects of the project: Kentaro Mizouchi and Miles Constable with the interface box to the T2K DAQ, Clive Mark, Mike Gallop and Chad Fisher with the remote handling design and testing, and Victor Verzilov with early design considerations.

This work has been supported by  the following agencies: the Natural Sciences and Engineering Research Council (NSERC) of Canada; TRIUMF;  the National Research Council, Canada;  the JSPS Fellowship program.



\bibliographystyle{model1a-num-names}
\bibliography{otrnim}

\begin{thebibliography}{16}
\expandafter\ifx\csname natexlab\endcsname\relax\def\natexlab#1{#1}\fi
\providecommand{\bibinfo}[2]{#2}
\ifx\xfnm\relax \def\xfnm[#1]{\unskip,\space#1}\fi
\bibitem[{Abe et~al.(2011)}]{t2k_nim}
\bibinfo{author}{K.~Abe}, et~al., \bibinfo{journal}{Nucl. Instrum. Meth. A}
  \bibinfo{volume}{659} (\bibinfo{year}{2011}).
\bibitem[{Ginsburg and Frank(1946)}]{ginsburg}
\bibinfo{author}{V.~Ginsburg}, \bibinfo{author}{I.~Frank},
  \bibinfo{journal}{JETP} \bibinfo{volume}{16} (\bibinfo{year}{1946}).
\bibitem[{Goldsmith and Jelley(1959)}]{goldsmith}
\bibinfo{author}{P.~Goldsmith}, \bibinfo{author}{J.~V. Jelley},
  \bibinfo{journal}{Phil. Mag.} \bibinfo{volume}{4} (\bibinfo{year}{1959}).
\bibitem[{Scarpine et~al.(2004)}]{fnal_otr}
\bibinfo{author}{V.~Scarpine}, et~al., \bibinfo{journal}{IEEE Transactions on
  Nuclear Science} \bibinfo{volume}{51} (\bibinfo{year}{2004}).
\bibitem[{Bosser et~al.(1985)}]{cern_otr}
\bibinfo{author}{J.~Bosser}, et~al., \bibinfo{journal}{Nucl. Instrum. Methods
  A} \bibinfo{volume}{238} (\bibinfo{year}{1985}).
\bibitem[{Toyoda(2009)}]{jparc_otr}
\bibinfo{author}{A.~Toyoda}, in: \bibinfo{booktitle}{Proceedings of DIPAC09}.
\bibitem[{Jackson(1998)}]{jackson}
\bibinfo{author}{J.~D. Jackson}, \bibinfo{title}{Classical Electrodynamics},
  \bibinfo{publisher}{John Wiley and Sons}, \bibinfo{edition}{third} edition,
  \bibinfo{year}{1998}.
\bibitem[{Ter-Mikaelian(1972)}]{ter-mikaelian}
\bibinfo{author}{M.~L. Ter-Mikaelian}, \bibinfo{title}{High-energy
  Electromagnetic Processes in Condensed Media}, \bibinfo{publisher}{John Wiley
  and Sons}, \bibinfo{edition}{english language} edition, \bibinfo{year}{1972}.
\bibitem[{Gitter(1992)}]{otr_gitter}
\bibinfo{author}{B.~Gitter}, \bibinfo{title}{Optical Transition Radiation},
  \bibinfo{type}{Technical note} \bibinfo{number}{CAA-TECH-NOTE-internal-\#24},
  UCLA, \bibinfo{year}{1992}.
\bibitem[{Mokhov(1995)}]{MARS_1}
\bibinfo{author}{N.~V. Mokhov}, \bibinfo{title}{The Mars Code System User's
  Guide}, \bibinfo{type}{Technical Report} \bibinfo{number}{Fermilab-FN-628},
  Fermilab, \bibinfo{year}{1995}.
\bibitem[{Krivosheev and Mokhov(2000)}]{MARS_2}
\bibinfo{author}{O.~Krivosheev}, \bibinfo{author}{N.~Mokhov}, in:
  \bibinfo{booktitle}{Proc. Monte Carlo 2000 Conf.(also in
  Fermilab-Conf-00/181)}.
\bibitem[{Mokhov(2003)}]{MARS_3}
\bibinfo{author}{N.~V. Mokhov}, \bibinfo{title}{Status of MARS Code},
  \bibinfo{type}{Technical Report} \bibinfo{number}{Fermilab-Conf-03/053},
  Fermilab, \bibinfo{year}{2003}.
\bibitem[{Mokhov et~al.(2004)Mokhov, Gudima, James, and et~al.}]{MARS_4}
\bibinfo{author}{N.~V. Mokhov}, \bibinfo{author}{K.~K. Gudima},
  \bibinfo{author}{C.~C. James}, \bibinfo{author}{et~al.},
  \bibinfo{title}{Recent Enhancements to the MARS15 Code},
  \bibinfo{type}{Technical Report} \bibinfo{number}{Fermilab-Conf-04/053},
  Fermilab, \bibinfo{year}{2004}.
\bibitem[{Ritt et~al.(2001)Ritt, Amaudruz, and Olchanski}]{midas}
\bibinfo{author}{S.~Ritt}, \bibinfo{author}{P.~Amaudruz},
  \bibinfo{author}{K.~Olchanski}, \bibinfo{title}{{MIDAS (Maximum Integration
  Data Acquisition System)}},
  \bibinfo{howpublished}{\url{http://midas.psi.ch}}, \bibinfo{year}{2001}.
\bibitem[{McCarthy et~al.(2003)}]{fluorescence}
\bibinfo{author}{K.~J. McCarthy}, et~al., \bibinfo{journal}{Jour. Nucl. Mater.}
  \bibinfo{volume}{321} (\bibinfo{year}{2003}).
\bibitem[{Kubota et~al.(1968)Kubota, Takahashi, and Doke}]{helium}
\bibinfo{author}{S.~Kubota}, \bibinfo{author}{T.~Takahashi},
  \bibinfo{author}{T.~Doke}, \bibinfo{journal}{Phys. Rev.}
  \bibinfo{volume}{165} (\bibinfo{year}{1968}) \bibinfo{pages}{225--230}.

\end{thebibliography}

\end{document}